\documentclass[12pt]{article}
\usepackage{latexsym,mathabx}

\newcommand{\sect}[1]{\setcounter{equation}{0}\section{#1}}

\newcommand{\eq}{\begin{equation}}
\newcommand{\eqa}{\begin{eqnarray}}
\newcommand{\en}{\end{equation}}
\newcommand{\ena}{\end{eqnarray}}
\newcommand{\enn}{\nonumber \end{equation}}

\newcommand{\CCC}{C}

\def\sk{\vskip .4cm}
\def\noi{\noindent}
\def\om{\omega}

\def\ga{\gamma}

\let \si\sigma
\let \part\partial

\def\unquarto{{1 \over 4}}

\def\unmezzo{{1 \over 2}}
\def\epsi{\varepsilon}
\def\we{\wedge}

\def\de{\delta}

\def\part{\partial}

\def\sk{\vskip .4cm}

\def\noi{\noindent}

\def\X0{X^0}

\def\om{\omega}

\def\ga{\gamma}

\def\unquarto{{1 \over 4}}
\def\unmezzo{{1 \over 2}}
\def\epsi{\varepsilon}

\def\we{\wedge}

\def\de{\delta}

\def\L#1#2{ \La^{#1}_{~~~#2} }

\def\square{{\,\lower0.9pt\vbox{\hrule \hbox{\vrule height 0.2 cm
\hskip 0.2 cm \vrule height 0.2 cm}\hrule}\,}}

\def\Lcal{{\cal L}}

\def\Phi{\phi}

\def\westar{\we_\star}

\def\omtilde{\tilde \om}
\def\Vtilde{\tilde V}

\def\Rtilde{\tilde R}
\def\rtilde{\tilde r}
\def\epsitilde{\tilde \epsi}

\def\psibar{\bar \psi}

\def\Om{\Omega}
\def\ppsi{fermionic}

\def\Ahat{{\hat A}}
\def\epsihat{{\hat \epsi}}
\def\dehat{{\hat \de}}
\def\Fhat{{\hat F}}
\def\fhat{{\hat f}}
\def\ghat{{\hat g}}
\def\phihat{{\hat \phi}}

\def\L{{\epsi}}
\def\*{\star}
\def\m*{\star^{op}}

\begin{document}

\begin{titlepage}
\begin{center}{\Large \bf Noncommutative gravity coupled to fermions: \\ second order expansion via Seiberg-Witten map}
\\[3em]
{\large {\bf Paolo Aschieri} and {\bf Leonardo Castellani}} \\ [2em] {\sl Dipartimento di Scienze e Tecnologie Avanzate and
\\ INFN Gruppo collegato di Alessandria,\\Universit\`a del Piemonte Orientale,\\ Viale T. Michel 11,  15121 Alessandria, Italy}\\ [1.5em]
\end{center}

\begin{abstract}

\vskip 0.2cm
We use the Seiberg-Witten map (SW map) to expand noncommutative gravity coupled to fermions
in terms of ordinary commuting fields. The action is invariant under general coordinate transformations and
local Lorentz rotations, and  has the same degrees of freedom as the commutative gravity action.
The expansion is given up to second order in the noncommutativity parameter $\theta$.

A geometric reformulation and generalization of the SW map is presented
that applies to any abelian twist. Compatibility of the map with
hermiticity and charge conjugation conditions is proven. The action 
is shown to be real and  invariant under charge
conjugation at
all orders in $\theta$.  This implies the bosonic part of the action to be even
in $\theta$, while the fermionic part is even in $\theta$ for Majorana fermions.
\end{abstract}

\vskip 9cm \noi \hrule \vskip.2cm \noi {\small aschieri@to.infn.it\\
leonardo.castellani@mfn.unipmn.it }

\end{titlepage}

\newpage
\setcounter{page}{1}

\sect{Introduction}

Field theories defined on noncommutative spaces can be
systematically constructed by use of an associative and
noncommutative  $\star$-product. This product between fields
generates infinitely many derivatives and introduces a
dimensionful noncommutativity parameter $\theta$.  The
prototypical and simplest example of $\star$-product is the
Groenewold-Moyal product \cite{MoyalGroenewold} (historically
arising in phase-space after Weyl quantization \cite{Weyl}) : 

\eqa  f(x) \star g(x) &\!\!\equiv\!\!& \exp
\left(  {i \over 2} \theta^{\mu\nu}  {\part \over \part x^\mu}
{\part \over \part y^\nu} \right) f(x) g(y) |_{y \rightarrow x} \label{starproduct} \\
 & \!\!= \!\!& f(x) g(x) + {i\over 2} \theta^{\mu \nu}
 \part_\mu f  \part_\nu  g + {1 \over 2!}  {\left( i \over 2 \right)^2} \theta^{\mu_1\nu_1} \theta^{\mu_2\nu_2} (\part_{\mu_1}   \part_{\mu_2} f )(\part_{\nu_1}   \part_{\nu_2} g )+ \cdots \nonumber
 \ena
 \noi with a constant $\theta$.

Replacing in the classical theory the usual product by the star
product leads to a deformation of the classical theory 
(called the noncommutative theory, or NC theory),
containing an infinite number of new interactions and higher derivative
terms. This procedure has been exploited to obtain deformed
gravity theories in various dimensions
\cite{Banados,MilanogroupG2,MilanogroupG3,Cardella,Chamseddine,Wessgroup,Estrada,AC1},
including deformed supergravity \cite{MilanogroupSG3,AC2}.

Such theories, seen as effective field theories,  encode {\sl ab initio}
a noncommutative structure of spacetime, and it may be interesting to
compare them to effective field theories emerging for example from 
string/brane interactions.

NC field theories  are invariant under deformations of the classical symmetries:
for instance the NC action for gauge fields is invariant under 
deformed gauge symmetries that involve $\star$ - products. 

The fields of NC theories can be expanded in a formal series in $\theta$. 
Via the Seiberg-Witten (SW) map \cite{SW}, this expansion can be
realized in terms of classical fields  (i.e. the original fields
of the ordinary theory) transforming under the ordinary transformation laws  \cite{Madore,Jurco1,Jurco2}. 
In fact  the SW map is explicitly
determined by requiring that the NC transformation laws of the NC fields arise from
the ordinary transformations of the classical fields, once the NC
fields are seen as function of the classical fields. The SW map is a key ingredient in the
construction of the NC standard model \cite{SM, AJSW}, because it allows to
have the same gauge group and degrees of freedom as in the commutative
case.

In this paper we apply the Seiberg-Witten map to the NC vielbein gravity theory
coupled to fermions developed in Ref.  \cite{AC1}. We determine the NC
vielbein, spin connection, curvature and
fermionic matter up to second order in $\theta$, and expand the NC action
to second order. We thus extend ordinary (commutative) vielbein gravity via
higher order derivative terms dictated by NC vielbein gravity. All these results are 
given first for the usual Groenewold-Moyal product, and then for a general
abelian twist determined by a set of commuting vector fields $\{X_I\}$.

The resulting action is a deformation of the commutative action,  given in terms
of the ordinary gravity fields and of the background noncommutativity vector
fields $\{X_I\}$. 
All fields transform covariantly  under ordinary diffeomorphisms, 
including the backgound fields $\{X_I\}$. These fields can be given dynamics by adding
covariant kinetic and potential terms, in the spirit of ref. \cite{ACD}.

The action, being written geometrically,  is invariant under general coordinate transformations.

Finally the expanded action is invariant under ordinary local Lorentz transformations
(gauge trasnformations) of the classical fields, since these classical
transformations induce, via the SW map, 
the NC transformations of the NC fields under which the NC action is
invariant. Thus local Lorentz invariance is not broken in the
$\theta$-expanded NC action: in fact  each order in $\theta$ of the
expanded action is separately gauge invariant.
This is not obvious at first sight because noncovariant terms, containing for example the``naked" spin connection, enter the expansion of the NC fields via the SW map. 
Only repeated  integrations by parts allow to
re-express the action  in terms of  gauge covariant quantities, where the
spin connection appears in Lorentz covariant derivatives and curvatures.

\sk

The organization of the paper is as follows. In Section 2 we begin with a
summary of NC vielbein gravity coupled to fermions and then prove
the charge conjugation invariance of the action.  In Section 3 the
Seiberg-Witten map at all orders is recalled, and in Section 4 the map
is found in the geometric language of exterior forms for a general abelian twist.
In Section 5 all the fields of the NC theory are expressed via the SW
map, up to second order in $\theta$, in terms of the classical
vielbein, spin connection, and Dirac fermions. In Section 6 we prove
that the SW map is compatible 
 with the hermiticity and charge conjugation conditions, ensuring that
 the action is real and even in $\theta$. In Section 7  the action  is expanded to second order. 
Section 8 contains our conclusions. 

Appendix A is devoted to a short summary of twisted differential
geometry, Appendix B deals with the hermiticity and charge conjugation
of the SW map in a general setting. Appendix C contains gamma matrix conventions and properties.

\sect{Noncommutative vielbein gravity coupled to fermions}

\subsection{Classical action}

The usual action of first-order gravity coupled to fermions can be recast
in an index-free form, convenient for generalization to the
non-commutative case:

\eq S =  \int Tr \left(i R \we V \we V \ga_5- [(D\psi) \psibar -
\psi D\psibar] \we V \we V \we V \ga_5 \right) \label{action1}
\en

\noi The fundamental fields are the 1-forms $\Om$  (spin
connection), $V$ (vielbein) and the fermionic 0-form $\psi$ (spin
1/2 field). The curvature 2-form $R$ and the exterior covariant
derivative on $\psi$ and $\psibar$ are defined by
 \eq
  R= d\Om - \Om \we
\Om, ~~~~~ D\psi = d\psi - \Om \psi,~~~~~D \psibar = d \psibar + \psibar  \Om    \label{psipsi}
\en
\noi with \eq \Om = {1 \over 4} \om^{ab} \ga_{ab}, ~~~~~V = V^a
\ga_a ~~~~~~
\en
\noi and thus are $4 \times 4$ matrices in the spinor representation.
 See Appendix C for $D=4$ gamma matrix conventions and useful relations.
  The Dirac conjugate is
defined as usual: $\psibar = \psi^\dagger \ga_0$.
 Then also $(D\psi) \psibar$, $\psi D\psibar$ are
matrices in the spinor representation, and the trace
 $Tr$ is taken on this representation.
Using the $D=4$ gamma matrix  identities:
\eq
\ga_{abc} = i \epsi_{abcd} \ga^d \ga_5,~~~~~~~
 Tr (\ga_{ab} \ga_c \ga_d \ga_5) = -4 i \epsi_{abcd}
  \en
 \noi leads to the usual action:
   \eq
   S = \int  R^{ab} \we V^c \we V^d   \epsi_{abcd} + i [ \psibar  \ga^a D\psi -
   (D \psibar)  \ga^a \psi ] \we V^b \we V^c \we V^d  \epsi_{abcd}
   \label{action1comp}
\en
\noi with
\eq
  R \equiv {1\over4} R^{ab} \ga_{ab},~~~R^{ab} = d\om^{ab}  - \om^a_{~c} \we \om^{cb}
\en

\subsection{Invariances}

The action is invariant under local diffeomorphisms
 (it is the integral of a 4-form on a 4-manifold)
  and under the local Lorentz rotations:
\eq
\de_\epsilon V = -[V,\epsilon ] , ~~~\de_\epsilon \Om = d\epsilon - [\Om,\epsilon],~~~~
\de_\epsilon \psi = \epsilon \psi, ~~~\de_\epsilon \psibar = -\psibar \epsilon
\en
\noi with
\eq
 \epsilon = {1\over 4} \epsilon^{ab} \ga_{ab}
  \en
  The invariance can be directly checked on the action (\ref{action1}) noting that
   \eq
  \de_\epsilon  R = - [{ R},\epsilon ]  ~~~\de_\epsilon D\psi = \epsilon D\psi,~~~\de_\epsilon ((D\psi) \psibar)
  = -[ (D\psi) \psibar, \epsilon], ~~~\de_\epsilon (\psi D\psibar) =
      -[\psi D \psibar, \epsilon],
   \en
   \noi using the cyclicity of the trace $Tr$ (on spinor indices) and the fact that $\epsilon$ commutes with
   $\ga_5$. The Lorentz rotations close on the Lie algebra:
   \eq
   [\de_{\epsilon_1},\de_{\epsilon_2}] = -\de_{[\epsilon_1,\epsilon_2]}
   \en

\subsection{Hermiticity and charge conjugation}

Since the vielbein $V^a$ and the spin connection $\om^{ab}$ are
real fields, the following conditions hold:

\eqa
 & & \ga_0 V \ga_0 = V^\dagger,~~~-\ga_0 \Omega \ga_0 =
 \Omega^\dagger, \\
 & &
  \ga_0 [(D\psi) \psibar] \ga_0 =  [\psi D \psibar]^\dagger,~~~
   \ga_0 [\psi D \psibar] \ga_0 =  [(D\psi)  \psibar]^\dagger
 \label{hermconj} \ena

\noi and can be used to check that the action (\ref{action1}) is
real.

 Moreover, if $C$ is the $D=4$ charge conjugation matrix
 (antisymmetric and squaring to $-1$), we have

\eq
 C V C = V^T,~~~C \Omega C = \Omega^T \label{conjVOm}
 \en
 \noi since the matrices $C\ga_a$ and $C \ga_{ab}$ are symmetric.

 Similar relations hold for the gauge parameter $\epsilon= (1/4) \epsi^{ab} \ga_{ab}$:
 \eq
 - \ga_0 \epsilon \ga_0 =  \epsilon^\dagger, ~~~C \epsilon C = \epsilon^T
  \en
  \noi $\epsi^{ab}$ being real.

 The charge conjugation of fermions:
  \eq
  \psi^C \equiv C (\psibar)^T
  \en
  \noi can be extended to the bosonic fields $V$, $\Om$ 
  \eq
   V^C \equiv  C V^T C,~~~\Om^C \equiv C \Om^T C\label{CCF1CC}
   \en
   \noi Then the relations (\ref{conjVOm}) can be written as:
    \eq
     V^C = V,~~~\Om^C = \Om\label{CCF2CC}
     \en
     and are the analogues of the Majorana condition for the
     fermions:
    \eq
   \psi^C = \psi~~~\rightarrow \psibar = \psi^T C
   \en
    So far we have been treating $\psi$ as a Dirac fermion, and
    therefore reality of the action requires both terms in
    square brackets in the action (\ref{action1}) or
    (\ref{action1comp}).
    If $\psi$ is Majorana, the two terms give the same
    contribution, and only one of them is necessary.

\subsection{The noncommutative action and its invariances}
After  replacing
exterior products by deformed exterior products (see Appendix A on twist differential geometry), the action(\ref{action1})  becomes:

\eq
 S =  \int Tr \left(i {R}\westar V \westar V \ga_5 -[ (D\psi)
\star \psibar - \psi \star D\psibar] \westar V \westar V \westar V
\ga_5 \right) \label{action1NC}
\en

\noi with
 \eq R= d\Om - \Om \westar \Om, ~~~~~ D\psi = d\psi -
\Om \star \psi ~~~~~D \psibar = d \psibar + \psibar  \star \Om      \label{2.20}
\en

Almost all preceding formulae continue to hold, with
$\star$-products and $\star$-exterior products. However, the
expansion of the fundamental fields on the Dirac basis of gamma
matrices must now include new contributions\footnote{for example 
$\om^{ab} \ga_{ab} \westar \om^{cd} \ga_{cd} = \om^{ab} \westar \om^{cd} (-i \epsilon_{abcd} \ga^5 - 4 \de^a_c \ga_{bd} - 2 \de^{ab}_{cd} 1 )$ contains $1, \ga_5$ besides $\ga_{ab}$ matrices since the
$\westar$ product is not antisymmetric.}:
 \eq
  \Om = {1 \over 4} \om^{ab} \ga_{ab} + i \om 1 + \omtilde \ga_5, ~~~~~V = V^a
\ga_a + \Vtilde^a \ga_a \ga_5  ~~~~~~
\en
\noi Similarly for the curvature :
 \eq
 R= {1\over 4} R^{ab} \ga_{ab} + i r 1 + \rtilde \ga_5
  \en
 \noi and for the gauge parameter:
 \eq
 \epsilon = {1\over 4} \epsi^{ab} \ga_{ab} + i \epsi 1 + \epsitilde \ga_5
  \en
  \noi Indeed now the $\star$-gauge variations read:
  \eq
\de_\epsilon V = -V \star \epsilon + \epsilon \star V,
~~~\de_\epsilon \Om = d\epsilon - \Om \star \epsilon+ \epsilon
\star \Om,~~~~ \de_\epsilon \psi = \epsilon \star \psi,
~~~\de_\epsilon \psibar = -\psibar \star \epsilon
\label{stargauge}\en
 \noi and in the variations for $V$ and $\Om$ also anticommutators of gamma matrices appear,
 due to the noncommutativity of the $\star$-product. Since for example the anticommutator
 $\{ \ga_{ab},\ga_{cd} \}$ contains $1$ and $\ga_5$, we see that the corresponding fields
 must be included in the expansion of $\Om$. Similarly, $V$ must contain a $\ga_a \ga_5$ term due
 to $\{ \ga_{ab},\ga_{c} \}$. Finally, the composition law for gauge parameters becomes:
 \eq
   [\de_{\epsilon_1},\de_{\epsilon_2}] = \de_{\epsilon_2 \star \epsilon_1 -
   \epsilon_1 \star \epsilon_2 }
   \en
   \noi so that $\epsilon$ must contain the $1$ and $\ga_5$ terms, since they appear in the
   composite parameter $\epsilon_2 \star \epsilon_1 - \epsilon_1 \star \epsilon_2$.

   The invariance of the noncommutative action (\ref{action1NC}) under the $\star$-variations is
   demonstrated  in exactly the same way as for the commutative case, noting that
   \eq
  \de_\epsilon R = - R \star \epsilon+ \epsilon \star R, ~~~\de_\epsilon D\psi = \epsilon \star D\psi,~~~\de_\epsilon ((D\psi) \star \psibar) = - (D\psi) \star \psibar \star \epsilon + \epsilon \star (D\psi) \star \psibar
   \en
   \noi etc., and using now the fact that
   $\epsilon$ still commutes with $\ga_5$, and the cyclicity of the
   trace $Tr$ with respect to pointwise matrix products  and the graded cyclicity 
   of the integral with respect to the $\star$-produc, so that 
$\int
   Tr$ is graded cyclic.
   
   The local $\star$-symmetry satisfies the Lie algebra of $GL(2,C)$, and centrally extends 
   the $SO(1,3)$ Lie algebra of the commutative theory.
   
   Finally, the $\star$-action (\ref{action1NC}) is invariant under diffeomorphisms
   generated by the Lie derivative, in the sense that 
   \eq
    \int \Lcal_v ({\rm 4{-}form}) = \int (i_v d + d i_v) ({\rm 4{-}form}) = \int d (i_v ({\rm 4{-}form}))= boundary~ term
    \en
    since $d({\rm 4{-}form}) =0$ on a 4-dimensional manifold. In fact
    the action is geometrical (it is the integral of a $4$-form) and
    as such it is invariant under usual coordinate transformations.         

\subsection{Hermiticity and charge conjugation}

Hermiticity conditions can be imposed on $V$, $\Om$ and the gauge parameter $\epsilon$:
\eq
 \ga_0 V \ga_0 = V^\dagger,~~~ -\ga_0 \Omega \ga_0 =
 \Omega^\dagger,~~~ -\ga_0 \epsilon \ga_0 =
 \epsilon^\dagger \label{hermconjNC}
 \en
 \noi Moreover it is easy to verify the analogues of conditions
 (\ref{hermconj}):
 \eq
 \ga_0 [(D\psi) \star \psibar] \ga_0 = [\psi \star D\psibar]^\dagger,~~ \ga_0 [\psi \star D\psibar] \ga_0 
 = [D\psi \star \psibar]^\dagger \label{hermconjpsi}
\en
\noi These hermiticity conditions are consistent with the gauge variations, as in the
commutative case, and can be used to check that the action (\ref{action1NC}) is
real. On the component fields $V^a$, $\Vtilde^a$, $\om^{ab}$, $\om$, and $\omtilde$, and
on the component gauge parameters $\epsi^{ab}$, $\epsi$, and $\epsitilde$ the hermiticity conditions (\ref{hermconjNC}) imply that they are real fields.

The charge conjugation relations (\ref{conjVOm}), however, cannot be exported
to the noncommutative case as they are. Indeed they would imply the vanishing of the
component fields $\Vtilde^a$, $\om$, and $\omtilde$ (whose presence is necessary in the noncommutative case) and anyhow would not be consistent with the $\star$-gauge variations.

 An essential modification is needed, and makes
 use of the $\theta$ dependence of the noncommutative fields. This dependence
 will be made explicit in Section 3, using the Seiberg-Witten map. 
 At this stage we just assume that there is such a dependence. Then we can
 impose consistent charge conjugation conditions as follows:
 \eq
 C V_\theta (x) C = V_{-\theta} (x)^T,~~~C \Omega_\theta (x) C = \Omega_{-\theta} (x)^T,~~~
 C \epsi_\theta (x) C = \epsi_{-\theta} (x)^T \label{ccc}
 \en
These conditions can be checked to be consistent with the
$\star$-gauge transformations. For example $ C V_\theta (x)^T C$ can
be shown to transform in the same way as $V_{-\theta} (x)$:
\eqa
\de_\epsilon (C  V_\theta^T C) &= & C (\de_\epsilon V_\theta)^T C = 
C (- \epsilon_\theta^T \star_{-\theta} V_\theta^T
+ V_\theta^T \star_{-\theta} \epsilon_\theta^T )C= \nonumber \\ & =&\epsilon_{-\theta} \star_{-\theta} 
V_{-\theta}
- V_{-\theta} \star_{-\theta} \epsilon_{-\theta}= \de_\epsilon V_{-\theta}
\ena
 where we have used $C^2 = -1$ and the fact that the transposition of a $\star$-product
 of matrix-valued fields interchanges the order of the matrices but not of the $\star$-multiplied
 fields. To interchange both it is necessary to use the ``reflected'' $\star_{-\theta}$ product obtained
 by changing the sign of $\theta$, since
 \eq
  f \star_\theta g = g \star_{-\theta} f
  \en
  \noi for any two functions $f,g$.

\noi For the component fields and  gauge parameters the charge
conjugation conditions imply:
 \eqa & & V^a_\theta=V^a_{-\theta}, ~~~
\om^{ab}_\theta = \om^{ab}_{-\theta} \\ & & \Vtilde^a_\theta
=-\Vtilde^a_{-\theta}, ~~~ \om^{}_\theta=- \om^{}_{-\theta} ,~~~\omtilde^{}_\theta=
- \omtilde^{}_{-\theta}, \label{cconjonfields}
 \ena
 \noi Similarly
for the gauge parameters:
 \eqa & & \epsi^{ab}_\theta= \epsi^{ab}_{-\theta}
 \\ & &  \epsi^{}_\theta =- \epsi^{}_{-\theta} ,~~~\epsitilde^{}_\theta=-
\epsitilde^{}_{-\theta} \label{cconjonparam}
 \ena

Finally, let us consider the charge conjugate spinor:
 \eq
 \psi^C \equiv C (\psibar)^T 
 \en
 \noi It transforms under $\star$-gauge variations as:
  \eq
   \de_\epsilon \psi^C = C (\de_\epsilon \psibar)^T = C (-\psibar
   \star \epsilon)^T = C (- \epsilon^T \star_{-\theta} \psi^*)=C
   \epsilon^T C \star_{-\theta} C \psi^* = \epsilon^{}_{-\theta} \star_{-\theta}
   \psi^C
   \en
 \noi i.e. it transforms in the same way as $\psi_{-\theta}$. Then
 we can impose the noncommutative Majorana condition:
  \eq
   \psi^C_\theta= \psi^{}_{-\theta}~~ \Rightarrow ~~\psi^\dagger_\theta \ga_0
   = \psi^T_{-\theta}C
   \en
If the NC Majorana condition holds for $\psi$, it is immediate to verify that
 \eq
  C (D\psi_\theta \star_\theta \psibar_\theta) C =- (\psi \star D\psibar)^T_{-\theta} \label{Majorana}
  \en
  \noi in close analogy with the charge conjugation conditions (\ref{ccc}).

\subsection{Reality and charge conjugation invariance of the action}

Reality of the noncommutative action is proven by using the
hermiticity conditions on $V$, $\Om$, $R$ and on the fermion bilinears
$D\psi \star \psibar$ and $\psi \star D \psibar$ when comparing the
action (\ref{action1NC}) with its complex conjugate, obtained by
taking the Hermitian conjugate of the  4-form in the overall trace inside the integral.
\sk

We define noncommutative charge conjugation to be the following
transformation (extended linearly and multiplicatively to products of fields):
\eq\label{defCconj}
\psi\to\psi^{\;C}=C(\bar\psi)^T=-\gamma_0 C \psi^\ast\!~,~~
V\to V^{\,C}\equiv
C{\,V}^{\;T}C\!~,~~\Omega\to \Omega^{\,C}\equiv 
C{\:\!
\Omega_{}}^{\;T}C\!~,~~\star_\theta\to\star_\theta^C=\star_{-\theta}~
\!~,
\en
and consequently $\wedge_{\star_\theta}\to\wedge_{\star_\theta}^{\,C}=\wedge_{\star_{-\theta}}~.$ 
Then the action (\ref{action1NC}) is invariant under charge conjugation.
Indeed (setting for short $\wedge_{-\theta}\equiv\wedge_{\star_{-\theta}}$), 
\eqa
S^C \!\!\!&=\!\!\!&    \int Tr \left(i {R^C}\wedge_{-\theta} V^C \wedge_{-\theta} V^C \ga_5 -[ (D\psi^C)
\star_{-\theta} \psibar^C - \psi^C \star_{-\theta} D\psibar^C] \wedge_{-\theta} V \wedge_{-\theta} V \wedge_{-\theta} V
\ga_5 \right) \nonumber\\
&=\!\!\!&    \int Tr \left(i {R^C}\wedge_{-\theta} V^C \wedge_{-\theta} V^C \ga_5 -[ (D\psi^C)
\star_{-\theta} \psibar^C - \psi^C \star_{-\theta} D\psibar^C] \wedge_{-\theta} V \wedge_{-\theta} V \wedge_{-\theta} V
\ga_5 \right)^T \nonumber\\
&=\!\!\!&S
\ena
It may be useful to exhibit the various steps. Let us first concentrate on the bosonic 
part of the action. Then:
\eqa S^C_{bosonic}\!\!&=& i  \int Tr ( {R^C}\wedge_{-\theta} V^C
\wedge_{-\theta} V^C \ga_5 )^T =
- i  \int Tr ( {R^T}\wedge_{-\theta} V^T \wedge_{-\theta} V^T
C\ga_5C^{-1} )^T \nonumber\\
&=& -i \int Tr 
 \left( (V^T \we_{-\theta} V^T \ga_5^T)^T \we_\star R \right)  
= -i \int Tr \left( -(V^T \ga^T_5)^T \we_\star V \we_\star R \right) \nonumber \\ 
& =& i \int Tr ( \ga_5
 V \we_\star  V  \we_\star R) = i \int Tr (R\we_\star  \ga_5
 V \we_\star  V) 
 =   i \int Tr (R\we_\star  
 V \we_\star  V \ga_5) \nonumber \\ 
&=&S_{bosonic}
\ena
Similarly the fermionic part of the action satisfies
$S_{\ppsi}^C=S_{\ppsi}$.  Let's first consider the connection
terms in
\eqa
S_{\ppsi}&=&\int -\bar\psi \star V\wedge_\star V\wedge_\star V\wedge_\star
\gamma_5 d \psi- d\bar\psi \wedge_\star V\wedge_\star V\wedge_\star
V\star\gamma_5 \psi\nonumber\\ &&~~+
\bar\psi\star  V\wedge_\star V\wedge_\star
V\wedge_\star\gamma_5 \Omega\star \psi -
\bar\psi \star\Omega\wedge_\star V\wedge_\star V\wedge_\star
V\star\gamma_5 \psi~ ~~~~~~~
\ena
We find 
\eqa
(\bar\psi \star V\wedge_\star V\wedge_\star
V\wedge_\star\gamma_5 \Omega\star\psi)^C&=&
\overline{\psi^C} \star_{-\theta}V^C\wedge_{-\theta} V^C\wedge_{-\theta}
V^C\wedge_{-\theta}\gamma_5 \Omega^C\star_{-\theta}\psi^C \nonumber\\
&=& 
-{(\bar\psi)^T} \star_{-\theta} V^T\wedge_{-\theta} V^T\wedge_{-\theta}
V^T\wedge_{-\theta}\gamma^T_5 \Omega^T\star(\bar\psi)^T
\nonumber\\
&=& {\bar\psi}\star \Omega\gamma_5\wedge_\star V\wedge_\star
V\wedge_\star V\star\psi \nonumber\\
&=&-{\bar\psi}\star \Omega\wedge_\star V\wedge_\star
V\wedge_\star V\star\gamma_5\psi 
\ena
where in the third line we inserted the definitions of the charge conjugate
fields and simplified the $C$ matrices by recalling that
$\gamma_5^T=C\gamma_5C^{-1}$; in the fourth line we transposed
the whole expression (that is invariant because it is valued in
complex numbers).
We observe  that the charge conjugation transformation squares to the
identity; then the two connection terms  in $S_{\ppsi}$ are
mapped one into the other under charge conjugation, and hence their sum  
is invariant.  

The proof for the fermion kinetic term is similar, and can be obtained by
replacing the spin connection with the exterior derivative, 
invariant under charge conjugation.
\sk

We can also conclude that the action {\it must be even in} $\theta$ if
the fermions satisfy the Majorana condition.
Indeed  (\ref{ccc}) and  the Majorana fermions property (\ref{Majorana})  imply
\eq
V^{\,C}= 
V_{-\theta}~,~~
\Omega^{\,C}=\Omega_{-\theta}~,~~\star_\theta^C=\star_{-\theta}~,~~
[ (D\psi)
\star \psibar - \psi \star D\psibar]^C=
[ (D\psi)
\star \psibar - \psi \star D\psibar]_{-\theta}
\!~.
\en
Hence the bosonic action $S_{bosonic} (\theta) $ is mapped into
$S_{bosonic} (-\theta)$ under charge conjugation. Also for the fermionic
action $S_{fermionic} (\theta)$ we have $S_{fermionic}
(\theta)^C=S_{fermionic} (-\theta)$ 
if the fermions are Majorana. Invariance of  $S_{bosonic}$ and of
$S_{fermionic}$ under charge conjugation then implies invariance of
the action under $\theta\to -\theta$.
Finally $S(\theta)=S(-\theta)$ implies that all corrections to the
classical action are even in $\theta$ if we consider Majorana
fermions.

\subsection{Commutative limit $\theta \rightarrow 0$}

In the commutative limit the action reduces to the usual action of gravity coupled to fermions
of eq. (\ref{action1}). Indeed in virtue of the charge conjugation conditions on $V$ and $\Om$,
the component fields  $\Vtilde^a$, $\om$, and $\omtilde$ all vanish in the limit $\theta \rightarrow 0$
(see the second line of (\ref{cconjonfields})), and only the classical spin connection
$\om^{ab}$, vielbein $V^a$ and Dirac fermion $\psi$ survive. Similarly the gauge parameters
$\epsi$, and $\epsitilde$ vanish in the commutative limit.

\section{SW map for Groenewold-Moyal noncommutativity}
In this section we consider Groenewold-Moyal noncommutativity, i.e. the star
product is given by  (\ref{starproduct}).
The Seiberg-Witten map (SW map) relates the noncommutative gauge
fields $\Ahat$ to the ordinary $A$, and the noncommutative gauge
parameters $\epsihat$ to  the ordinary $\epsi$ and $A$ so as to satisfy:
 \eq
 \Ahat (A) + \dehat_\epsihat \Ahat (A) = \Ahat (A + \de_\epsi A) \label{SWcondition}
 \en
 with 
  \eqa 
   & &
  \de_\epsi A_\mu = \part_\mu \epsi + i \epsi A_\mu - i A_\mu 
      \epsi, \\
      & &
  \dehat_\epsihat \Ahat_\mu = \part_\mu \epsihat + i \epsihat \star \Ahat_\mu - i \Ahat_\mu \star     
      \epsihat;\label{3.44h}
      \ena
here, as usual in the literature on the subject,  the $A$ and 
$\epsi$  transformations are chosen to be compatible with hermiticity
(rather than anti-hermiticity) conditions.

The Seiberg-Witten condition (\ref{SWcondition}) states that the dependence of the noncommutative gauge field on the ordinary one is fixed
by requiring that ordinary gauge variations of $A$ inside $\Ahat(A)$ produce the noncommutative
gauge variation of $\Ahat$.  In a gauge theory physical quantities are
gauge invariant: they do not depend on the gauge potential but on
the equivalence class of potentials related by gauge transformations.
The SW map relates the noncommutative gauge theory to the
commutative one by requiring noncommutative fields  to have the same gauge equivalence
classes as the commutative ones. In this way the degrees of freedom of a
noncommutative gauge theory are the same as those of the corresponding
commutative one. 

Equation  (\ref{SWcondition}) can be solved order by order in $\theta$
\cite{Jurco2}, yielding
$\Ahat$ and $\epsihat$ as  power series in $\theta$:  
 \eqa 
    \Ahat (A, \theta) &=& A + A^1 (A)  + A^2 (A) + \cdots + A^n (A)+ \cdots  \\
     \epsihat (\epsi, A, \theta)  &=&  \epsi + \epsi^1 (\epsi, A)+ \epsi^2 (\epsi, A)+ \cdots + 
     \epsi^n (\epsi, A)+ \cdots 
     \ena    
 \noi  where $A^n (A)$ and $\epsi^n (\epsi, A)$  are of order $n$ in $\theta$. Note that  $\epsihat$
depends on the ordinary $\epsi$ and also on $A$.

In \cite{SW} it is shown that if $\Ahat$ and
$\epsihat$  solve the differential equations
   \eqa
 & & { \part \over \part \theta^{\rho\sigma}} \Ahat_\mu = - \unquarto \{ \Ahat_{[\rho}, \part_{\sigma ]} \Ahat_\mu + \Fhat_{\sigma]\mu} \}_\star\label{one} \\
 & &  { \part \over \part \theta^{\rho\sigma}} \epsihat =  - \unquarto \{  \Ahat_{[\rho}, \part_{\sigma]} \epsihat \}_\star\label{two}
 \ena
  with the definitions
  \eqa
  & &
   \Fhat_{\nu\rho} \equiv \part_\nu \Ahat_\rho -  \part_\rho \Ahat_\nu - i \Ahat_\nu \star \Ahat_\rho
    + i \Ahat_\rho \star \Ahat_\nu \label{Fhatmunu} \\
   & &
    \{f,g\}_\star \equiv f \star g + g \star f
    \ena
then $\Ahat$ and
$\epsihat$  satisfy also the SW condition (\ref{SWcondition}).

   The differential equations (\ref{one}),(\ref{two}) admit solutions given recursively by \cite{Ulker}
    \eqa
     & &
     A^{n+1}_\mu = -{1 \over 4(n+1)} \theta^{\rho\sigma} \{\Ahat_\rho, \part_\sigma \Ahat_\mu   
     + \Fhat_{\sigma\mu} \}^n_\star \label{rone}\\
      & & \epsi^{n+1}=  -{1 \over 4(n+1)} \theta^{\rho\sigma} \{\Ahat_\rho, \part_\sigma \epsihat \}^n_\star\label{rtwo}
       \ena
   \noi where  $ \{ \fhat, \ghat \}^n_\star $ is $n$-th  order term in $ \{ \fhat, \ghat \}_\star $, so that for example
     \eq
     \{\Ahat_\rho, \part_\sigma \epsihat \}^n_\star \equiv \sum_{r+s+t=n} (A^r_\rho \star^s \part_\sigma \epsi^t +
      \part_\sigma \epsi^t \star^s A^r_\rho )
       \en
    \noi and $\star^s$ indicates the $s$-th order term in the star
    product expansion (\ref{starproduct}). Here is a simple proof of
    (\ref{rone}), (\ref{rtwo}):
   multiplying the differential equations by $\theta^{\mu\nu}$ and analysing them
   order by order yields
      \eqa
  & & \theta^{\mu\nu}  { \part \over \part \theta^{\mu\nu}} A_\rho^{n+1} = (n+1) A_\rho^{n+1}= - \unquarto  \theta^{\mu\nu}  \{ \Ahat_{[\mu}, \part_{\nu ]} \Ahat_\rho + \Fhat_{\nu]\rho} \}_\star^n\\
  & &  \theta^{\mu\nu}  { \part \over \part \theta^{\mu\nu}} \epsi^{n+1}=  (n+1) \epsi^{n+1}= - \unquarto  \theta^{\mu\nu}  \{  \Ahat_{[\mu}, \part_{\nu]} \epsihat \}_\star^n
  \ena
\noi since $A_\rho^{n+1}$ and $\epsi^{n+1}$ are homogeneous functions of $\theta$ of order 
$n+1$. 

Similar expressions hold for the gauge field strength, and for matter fields $\phi$  transforming
in the fundamental or in the adjoint representation of the gauge group:

\eqa
 & & F^{n+1}_{\mu\nu}  = -{1 \over 4(n+1)} \theta^{\rho\sigma} \left( \{\Ahat_\rho, \part_\sigma \Fhat_{\mu\nu}   
     + D_\sigma \Fhat_{\mu\nu} \}^n_\star - 2\{\Fhat_{\mu\rho},\Fhat_{\nu\sigma} \}^n_\star \right) \label{Frec}\\
&  &
      \phi^{n+1}= -{1 \over 4(n+1)} \theta^{\rho\sigma} \left(\Ahat_\rho \star (\part_\sigma \phihat
     + D_\sigma \phihat) \right)^n, ~~\dehat_\epsihat \phihat = i \epsihat \star \phihat \label{phirec} \\
 & & \phi^{n+1}= -{1 \over 4(n+1)} \theta^{\rho\sigma} \{\Ahat_\rho, \part_\sigma \phihat
     + D_\sigma \phihat \}^n_\star, ~~\dehat_\epsihat \phihat = i \epsihat \star \phihat - i \phihat \star \epsihat 
\label{phiadrec}
\ena
where the covariant derivative on $\Fhat$ and $\phihat$  is given by
$D_\sigma\hat{F}_{\mu\nu}=\partial_\sigma\hat{F}_{\mu\nu}-i\hat{A}_\sigma\star
\hat{F}_{\mu\nu}+i \hat{F}_{\mu\nu}\star\hat{A}_\sigma$,
$D_\sigma \phihat = \part_\sigma \phihat -i  \Ahat_\si \star \phihat $  (fundamental) and
$D_\sigma \phihat = \part_\sigma \phihat -i  \Ahat_\si \star \phihat + i \phihat \star \Ahat_\si$ (adjoint).

\sk
\noi {\bf Note} The solution to (\ref{SWcondition}) is not unique. For example
      if $\hat A$ is a solution, any finite noncommutative gauge
      transformation of $\hat A$ gives another solution. Another
      source of ambiguities is related to field redefinitons of the
      gauge potential.  Both types of ambiguities should not lead
      to physical effects  since the $S$ matrix
      of the $\theta$ expanded theory (the noncommutative theory) is
      order by order in $\theta$ gauge invariant and is expected to be
      independent from field redefinitions.  The ambiguities are anyhow 
      constrained by two physical requirements: 
\begin{itemize} 
\item{ the hermiticity
      properties of the commutative fields must be extended to the
noncommutative fields, and this implies reality of the NC actions;}
\item{
      the charge conjugation properties of the commutative fields must
      be extended to the noncommutative fields, and this implies that
      commutative actions with charge conjugation symmetry can be
      deformed into noncommutative ones with noncommutative charge
      conjugation symmetry. }
\end{itemize}
\sk
      In this respect  it is worth mentioning that the physically relevant
      ambiguities found up to second order in $\theta$ in
      \cite{Ohl}  do not preserve the charge conjugation properties of
      the noncommutative fields \cite{ACpreparation}.

It is therefore possible  that the expansion  of 
NC vielbein gravity to ${\cal O}(\theta^2)$   presented in the next Sections  is unique up to
physically irrelevant field redefinitions.

\section{Geometric formulation of SW map for a general abelian twist}

The SW map conditions can be  formulated  in the
presence of an arbitrary (space-time dependent) NC star-product, 
not necessarily only in the Groenewold-Moyal case. The  SW map with
arbitrary noncommutativity has been constructed for abelian gauge
theories in \cite{JSW} using Kontsevich's results. In the nonabelian case the
situation is more involved and there is no definite result (despite
 interesing partial ones \cite{JS}). In this section we show that
when the star-product is obtained via a set of mutually commuting vector
fields (i.e. via an abelian twist) we can construct order by order solutions to the SW map.
The mutually commuting vector fields can be spacetime dependent and hence
we obtain a SW map for nonabelian gauge fields with nonconstant
noncommutativity. The resulting NC gauge potential is then
a $1$-form that depends on the commutative gauge potential $1$-form
and on the mutually  commutative vector fields.

The SW condition (\ref{SWcondition}) can be seen as an $1$-form equation, and
therefore is coordinate independent.  Likewise the differential eq.s (\ref{one}),(\ref{two}) can be
recast in a coordinate independent form:
  \eqa
 & & { \part \over \part \theta^{IJ}} \Ahat= - \unquarto \{ i_{X_{[I}} \Ahat, \Lcal_{X_{J]}} \Ahat+  i_{X_{J]}} \Fhat \}_\star\label{onegeom} \\
 & &  { \part \over \part \theta^{IJ}} \epsihat =  - \unquarto \{  i_{X_{[I}} \Ahat, \Lcal_{X_{J]}} \epsihat \}_\star\label{twogeom}
 \ena
\noi where all quantities and operations are diffeomorphic-convariant: $\Ahat$ is a one-form, $\Fhat \equiv d\Ahat - i \Ahat \westar \Ahat$ is a two-form,
$i_{X_I}$ and $ \Lcal_{X_I}$  are respectively the contraction and the Lie derivative along the mutually
commuting vector fields $X_I$. When the abelian twist reduces to the Moyal case of  the preceding Section, the curvature becomes
 $\Fhat = \unmezzo  \Fhat_{\mu\nu} dx^\mu \we dx^\nu$, where the $ \Fhat_{\mu\nu}$ components are given in
(\ref{Fhatmunu}). Note that in the Moyal case $dx^\mu \westar dx^\nu =dx^\mu \we dx^\nu$ since $\Lcal_{X_I} dx^\mu = d \Lcal_{X_I}x^\mu=0$.

Our strategy is the following: we first write down, in a coordinate independent way, 
candidate recursive solutions for the  SW condition (\ref{SWcondition}). Then we prove that in a particular coordinate system the
candidate solutions satisfy the SW condition, and therefore must satisfy it in any coordinate system.

The candidate recursive solutions are given by:
 \eqa
     & &
     A^{n+1} = -{1 \over 4(n+1)} \theta^{IJ} \{i_{X_I} \Ahat, \Lcal_{X_J} \Ahat
     + i_{X_J}  \Fhat \}^n_\star \label{Asol}\\
      & & \epsi^{n+1}=  -{1 \over 4(n+1)} \theta^{IJ} \{ i_{X_I} \Ahat, \Lcal_{X_J}  \epsihat \}^n_\star\label{epsisol}
       \ena

We choose now a particular coordinate system adapted to the mutually commuting vectors $X_I$, i.e. precisely
the coordinates $y^I$ such that $X_I = \part / \part y^I$. It is then immediate to verify that the candidate
solutions indeed reduce to the recursive solutions given in the preceding Section, where one just substitutes
the coordinates $x^\mu$ with $y^I$. For example $\Lcal_{X_I} (\Ahat_J dy^J) = \part \Ahat_J / \part y^I dy^J$, etc. Thus (\ref{Asol}) and (\ref{epsisol}) are {\it bona fide}  solutions
of the SW equations in a generic coordinate system. The same argument can be used to prove that (\ref{Asol}) and 
(\ref{epsisol}) are solutions of the differential equations (\ref{onegeom}) and (\ref{twogeom}) , and that
these latter imply the SW gauge condition (\ref{SWcondition}).

Similarly one proves  the generalization of eq.s (\ref{Frec})-(\ref{phiadrec}):
\eqa
   F^{n+1} &\!\!=\!\!& -{1 \over 4(n+1)} \theta^{IJ} \left( \{ i_{X_I} \Ahat, 2 \Lcal_{X_J}  \Fhat  - i [ i_{X_J} \Ahat, \Fhat ]\}^n_\star -  [i_{X_I} \Fhat ,i_{X_J} \Fhat ]^n_\star \right) \nonumber \\ \label{Frecgen}\\
      \phi^{n+1}&\!\!=\!\!& -{1 \over 4(n+1)} \theta^{IJ} \left( i_{X_I} \Ahat  \star (2 \Lcal_{X_J} \phihat
      -i (i_{X_J}    \Ahat) \star \phihat ) \right)^n, ~~\dehat_\epsihat \phihat = i \epsihat \star \phihat \label{phirecgen} \\
 \phi^{n+1}&\!\!=\!\!& -{1 \over 4(n+1)} \theta^{IJ} \{  i_{X_I} \Ahat \,, \,2\Lcal_{X_J} \phihat
      - i (i_{X_J}\Ahat) \star \phihat + i \phihat \star (i_{X_J}\Ahat)\}_\star^n, ~~\dehat_\epsihat \phihat = i \epsihat \star \phihat- i\phihat\star\epsihat\nonumber\\
\label{phiadrecgen}
\ena

\sk
\noi {\bf Note} In order to  prove that (\ref{Asol}) and
(\ref{epsisol}) satisfy the SW condition we have implicitly assumed
that we have four commuting vector fields
$\{X_I\}$ spanning 
(at each point of the four dimensional space-time manifold) 
the four dimensional tangent space-time. In this case we
can indeed locally find a coordinate system $\{y^J\}$ such that 
$X_I= \part / \part y^I$ (Frobenius theorem).  
This assumption can 
be relaxed: (\ref{Asol}) and
(\ref{epsisol}) satisfy the SW condition also in the case of
abelian twists $e^{-{{i}\over 2}
\theta^{IJ}X_I\otimes X_J}$ where the $N$
mutually commuting vector fields $\{X_I\}$  span a 
subspace of the four dimensional tangent space-time.
The proof is algebraic and follows the same steps as the original
proof given by Seiberg and Witten.
\sk

\sect{Expansion of fields to second order in $\theta$}

Here we apply the formulae of the preceding section to the gauge
fields, matter fields and gauge parameters of  noncommutative vielbein 
gravity coupled to fermions. The gauge field
is the spin connection $\Omega$, the matter fields are the vielbein
(since it transforms in the adjoint representation of the gauge group,
see  (\ref{stargauge})) and the fermi field (transforming in the
fundamental representation).  
Comparison of the matter fields gauge
transformations  $D\psi = d\psi - \Om \star \psi$ (defined in (\ref{2.20})) and 
$D_\sigma \phihat = \part_\sigma \phihat -i  \Ahat_\si \star \phihat $
(defined below (\ref{phiadrec})), show that the formulae of the preceding
section holds with $i\widehat{A}$ replaced by  $\Omega$. 

The solutions of  the SW map provide explicit  expressions for the fields in terms of $\theta$ 
and the ordinary fields. The expressions for $V$, $\Om$ and $R$  must have the all order gamma
matrix structure:
\eqa
& & V=V^a \ga_a + \Vtilde^{1~a} \ga_a \ga_5 + V^{2~a} \ga_a + \Vtilde^{3~a} \ga_a \ga_5+ \cdots \label{SWsolV}\\
& & \Om = \unquarto \om^{ab} \ga_{ab} + ( i \om^11 +  \omtilde^1 \ga_5) + \unquarto \om^{2~ab} \ga_{ab} + ( i \om^31 +  \omtilde^3 \ga_5) + \cdots \label{SWsolom} \\
& & R = \unquarto R^{ab} \ga_{ab} + ( i R^11 +  \Rtilde^1 \ga_5) + \unquarto R^{2~ab} \ga_{ab} + ( i R^31 +  \Rtilde^3 \ga_5) + \cdots  \label{SWsolR}
\ena
\noi
(with all component fields $V^a, \tilde V^{1~a}, V^{2~a},
\tilde V^{3~a}, \omega^{ab}, \omega^1, \tilde\omega^1, etc.$ real)  as
can be deduced from (\ref{rone}), (\ref{phiadrec}), (\ref{Frec}).
For example the first order term in $\Om$
contains an anticommutator of $\ga_{ab}$ matrices, yielding $1$ and $\ga_5$ matrices. 
The second order term contains anticommutators of $\ga_{ab}$ with $1$ and $\ga_5$,
or commutators of $\ga_{ab}$ with $\ga_{ab}$, yielding again $\ga_{ab}$ matrices, and so on.
The gamma matrix structure depends therefore on the parity of the order in $\theta$.
The corrections up to second order in $\theta$ in the above expansions, and for the fermi field, 
are given in the following Tables, for Groenewold-Moyal twist, and for general abelian twist.
The expansion up to first order appeared
in  \cite{Miao:2010kr}
(see also \cite{Aschieri:2011ih}). 
\sk\sk\sk\sk
\noi { \bf TABLE 1:  SW fields at second order, Groenewold-Moyal product}

 \sk
\noi{\bf Vielbein}
\eqa
 & & V^{1~a}_\mu = 0 \\
 & & \Vtilde^{1~a}_\mu=  {1 \over 4} \theta^{\rho\sigma} \om^{bc}_\rho \epsilon^{abcd}
       (\part_\sigma V^d_\mu - \unmezzo \om^{de}_\sigma V^e_\mu)
\ena
  \eqa
  V^{2~a}_\mu = -{1 \over 8} \theta^{\rho\sigma} \left[ 4 \om^1_\rho (\part_\sigma V^a_\mu
 - \unmezzo \om^{ab}_\sigma V^b_\mu)- \epsilon_{abcd} \om^{bc}_\rho   (\part_\sigma \Vtilde^{1~d}_\mu
 + \omtilde^1_\sigma V^d_\mu  - \unmezzo \om_\sigma^{de} \Vtilde^{1~e}_\mu )
 \right]  \nonumber\\
  + {1 \over 8} \theta^{\rho\sigma} \theta^{\nu\tau} \left[  \om^{bc}_\rho \part_\nu \om^{bc}_\sigma \part_\tau V^a_\mu + 2 \om^{bc}_\rho \part_\nu \om^{ca}_\sigma \part_\tau V^b_\mu - \part_\nu \om^{ab}_\rho \part_\tau (\part_\sigma V^b_\mu - \unmezzo \om^{bc}_\sigma V^c_\mu)  \right]\\
  \Vtilde^{2~a}_\mu = 0 ~~~~~~~~~~~~~~~~~~~~~~~~~~~~~~~~~~~~~~~~~~~~~~~~~~~~~~~~~~~~
  ~~~~~~~~~~~~~~~~~~~~~~~~~~~~~
  \ena
{\bf Spin connection}
\eqa
     & & \om^{1~ab}_\mu =0 \\
    & & \om^1_\mu = - {1 \over 16} \theta^{\rho\sigma} \om_{\rho}^{ab} (\part_\sigma \om^{ab}_\mu + R^{ab}_{\sigma\mu} ) \\
 & &  \omtilde^1_\mu = - {1 \over 16} \theta^{\rho\sigma} \om^{ab}_\rho (\part_\sigma \om^{cd}_\mu + R^{cd}_{\sigma\mu}) \epsilon_{abcd}  
 \ena
  \eqa
& &  \om^{2~ab}_\mu = - {1\over 8} \theta^{\rho\sigma} (\part_\sigma \om^{cd}_\mu + R^{cd}_{\sigma\mu}  ) (2 \om^1_\rho \de^{ab}_{cd} + \omtilde^1_\rho \epsilon_{abcd} )  \nonumber \\
 & & ~ ~ ~ ~ ~ ~ ~ ~ ~   - {1\over 4} \theta^{\rho\sigma}   \om^{cd}_\rho [(\part_\sigma
 \om^1_\mu  + R^1_{\sigma\mu} ) \de^{ab}_{cd} + \unmezzo (  \part_\sigma
 \omtilde^1_\mu  + \Rtilde^1_{\sigma\mu} ) \epsilon_{abcd}]  \nonumber \\
 & & ~ ~ ~ ~ ~ ~ ~ ~ ~ +  {1\over 8} \theta^{\rho\sigma} \theta^{\nu\tau} (\part_\nu \om^{ac}_\rho ) \part_\tau
 (\part_\sigma \om^{bc}_\mu + R^{bc}_{\sigma\mu}) \\
 & & \om^2_\mu = 0 \\
 & & \omtilde^2_\mu = 0
 \ena
 {\bf Curvature} 
 \eqa
 & &  R^{1~ab}_{\si\mu} = 0\\
& &  R^1_{\sigma\mu} = - {1 \over 16} \theta^{\nu\tau} [\om^{ab}_\nu (\part_\tau R^{ab}_{\sigma\mu} + D_\tau R^{ab}_{\sigma\mu}) - 2 R^{ab}_{\sigma\nu} R^{ab}_{\mu\tau}] \\
& & \Rtilde^1_{\sigma\mu} = {1 \over 32} \theta^{\nu\tau} [\om^{ab}_\nu (\part_\tau R^{cd}_{\sigma\mu} + D_\tau R^{cd}_{\sigma\mu}) \epsilon_{abcd}  - 2 R^{ab}_{\sigma\nu} R^{cd}_{\mu\tau} \epsilon_{abcd}] 
\ena
\eqa
& & R^{2~ab}_{\mu\nu} = -{1 \over 4} \theta^{\rho\si} [ 2 \om^{ab}_\rho \part_\si R^1_{\mu\nu} + 2 
 \omtilde^{ab}_\rho \part_\si \Rtilde^1_{\mu\nu}  \nonumber \\
 & & ~ ~ ~ ~ ~ ~ ~ ~ ~ ~ +  \om^1_\rho (\part_\si R^{ab}_{\mu\nu}+
 D_\si R^{ab}_{\mu\nu})+ \omtilde^1_\rho (\part_\si \Rtilde^{ab}_{\mu\nu}+
 D_\si \Rtilde^{ab}_{\mu\nu}) \nonumber \\
 & & ~ ~ ~ ~ ~ ~ ~ ~ ~ ~   - 2 R^{ab}_{\mu\rho} R^1_{\nu\si} - 2 \Rtilde^{ab}_{\mu\rho} R^1_{\nu\si} 
    - 2 R^1_{\mu\rho} R^{ab}_{\nu\si} - 2 \Rtilde^1_{\mu\rho} R^{ab}_{\nu\si}] \nonumber\\
    & & ~ ~ ~ ~ ~ ~ ~ ~ ~ ~ + {1 \over 32}  \theta^{\rho\si}   \theta^{\lambda\tau} [ \omtilde^{ab}_\lambda \part_\rho 
    \omtilde^{cd}_\tau \part_\si R^{cd}_{\mu\nu} -  \om^{ab}_\lambda \part_\rho 
    \om^{cd}_\tau \part_\si R^{cd}_{\mu\nu} \nonumber \\
    & & ~ ~ ~ ~ ~ ~ ~ ~ ~ ~ - 4 \part_\rho \om^{ca}_\lambda \part_\si (\part_\tau R^{cb}_{\mu\nu} + D_\tau R^{cb}_{\mu\nu} )+ 8 \part_\rho R^{ca}_{\mu\lambda} \part_\si R^{cb}_{\nu\tau}] \\
   & & R^2_{\mu\nu} = 0 \\
    & & \Rtilde^2_{\mu\nu} = 0
\ena
~ ~ ~ ~ ~ ~ ~ ~ ~ ~ with $\omtilde^{ab}_\mu \equiv \unmezzo \epsilon^{abcd} \om^{cd}_\mu$, 
$\Rtilde^{ab}_{\mu\nu} \equiv \unmezzo \epsilon^{abcd} R^{cd}_{\mu\nu}$.
\sk
\noi {\bf Fermion field}
\eqa
 \psi^1&=& {1 \over 8} \theta^{\mu\nu} \om^{ab}_\mu (\ga_{ab} \part_\nu \psi + \om^{ac}_\nu \ga_{bc} \psi) \\
\psi^2 &=& - {1\over 8} \theta^{\rho\sigma} [(\om^1_\rho - i \ga_5 \omtilde^1_\rho)(\part_\si \psi + D_\si \psi) -{i\over 4}
  \om^{ab}_\rho \ga_{ab} (\part_\si \psi^1 + D_\si \psi^1) - ( i \om^1_\si + \ga_5 \omtilde^1_\si ) \psi ] \nonumber \\
  & &   ~~~~~~~+ {1 \over 64} \theta^{\rho\si} \theta^{\mu\nu} \ga_{ab} [ {1 \over 4} \om^{ab}_\rho \ga_{cd} \part_\mu \om^{cd}_\si \part_\nu \psi
    +  \part_\mu \om^{ab}_\rho \part_\nu (\part_\si \psi + D_\si \psi)]
   \ena

\sk
\noi {\bf TABLE 2: SW forms at second order, general abelian twist}
 \sk
\noi {\bf Vielbein}
\eqa
V^{1~a} &=& 0 \\
\Vtilde^{1~a} &=&  {1 \over 4} \theta^{IJ}  X_{I}^{\rho} \om^{bc}_\rho \epsilon^{abcd}
       (\Lcal_{X_{J }}V^d - \unmezzo X_{J}^{\sigma}\om^{de}_\sigma V^e)
\ena
  \eqa
    V^{2\;a}\!\!\! &=& \!\!\!-{1 \over 8} \theta^{IJ} X_{I}^{\rho} \left[ 4 \om^1_\rho (\Lcal_{X_{J}}V^a
 - \unmezzo X^\sigma_J \om^{ab}_\sigma V^b)- \epsilon_{abcd} \om^{bc}_\rho   (\Lcal_{X_J} \Vtilde^{1~d} +X^\sigma_J \omtilde^1_\sigma V^d  - \unmezzo X^\sigma_J  \om_\sigma^{de} \Vtilde^{1~e} ) \right]  \nonumber\\
 & & ~~~~ + {1 \over 8} \theta^{IJ} \theta^{KL} [ X^\rho_I 
 \om^{bc}_\rho \Lcal_{X_K} (X^\sigma_J  \om^{bc}_\sigma ) 
\Lcal_{X_L} V^a +  2 \om^{bc}_\rho \Lcal_{X_K} (X^\sigma_J \om^{ca}_\sigma) 
 \Lcal_{X_L} V^b  \nonumber \\
& & ~~~~~~~~~~~~~    -\Lcal_{X_K} (X^\rho_I \om^{ab}_\rho ) \Lcal_{X_L} (\Lcal_{X_J} V^b - \unmezzo X^\sigma_J \om^{bc}_\sigma V^c) ] \\
\Vtilde^{2~a} &=&0 
  \ena
{\bf Spin connection}
\eqa
     & & \om^{1~ab}=0 \\
    & & \om^1= - {1 \over 16} \theta^{IJ} X_I^\rho \om_{\rho}^{ab} (\Lcal_{X_J} \om^{ab}+ i_{X_J} R^{ab} ) \\
 & &  \omtilde^1_\mu = - {1 \over 16} \theta^{IJ} X^\rho_I \om^{ab}_\rho (\Lcal_{X_J} \om^{cd} +
i_{X_J} R^{cd}) \epsilon_{abcd}  
 \ena
  \eqa
& &  \om^{2~ab} = - {1\over 8} \theta^{IJ} X^\rho_I  (\Lcal_{X_J} \om^{cd} + i_{X_J} R^{cd}  ) (2 \om^1_\rho \de^{ab}_{cd} + \omtilde^1_\rho \epsilon_{abcd} )  \nonumber \\
 & & ~ ~ ~ ~ ~ ~ ~ ~ ~   - {1\over 4} \theta^{IJ} X^\rho_I   \om^{cd} [(\Lcal_{X_J} 
 \om^1 +i_{X_J} R^1 ) \de^{ab}_{cd} + \unmezzo (  (\Lcal_{X_J} 
 \omtilde^1  +i_{X_J} \Rtilde^1 ) \epsilon_{abcd}]  \nonumber \\
 & & ~ ~ ~ ~ ~ ~ ~ ~ ~ +  {1\over 8} \theta^{IJ} \theta^{KL} \Lcal_{X_K} (X^\rho_I \om^{ac}_\rho ) 
 \Lcal_{X_L}
 (\Lcal_{X_J} \om^{bc}+ i_{X_J} R^{bc}) \\
 & & \om^2 = 0 \\
 & & \omtilde^2= 0
 \ena
 {\bf Curvature}
 \eqa
 & &  R^{1~ab} = 0\\
& &  R^1= - {1 \over 16} \theta^{IJ} [X^\nu_I \om^{ab}_\nu (2\Lcal_{X_J} R^{ab}- 2X^\tau_J \om^{ac}_\tau R^{cb}) - 2 (i_{X_I} R^{ab})( i_{X_J}  R^{ab})] \\
& & \Rtilde^1=  {1 \over 32} \theta^{IJ} [X^\nu_I \om^{ab}_\nu (2\Lcal_{X_J} R^{cd}- 2X^\tau_J \om^{ce}_\tau R^{ed}) \epsilon_{abcd} - 2 (i_{X_I} R^{ab})( i_{X_J}  R^{cd}) \epsilon_{abcd}]\nonumber\\
\ena
\eqa
& & R^{2~ab} = -{1 \over 2} \theta^{IJ} [  X^\rho_I \om^{ab}_\rho \Lcal_{X_J} R^1 +  
 X^\rho_I \omtilde^{ab}_\rho \Lcal_{X_J} \Rtilde^1  \nonumber \\
 & & ~ ~ ~ ~ ~ ~ ~ ~ ~ ~ + X^\rho_I  \om^1_\rho ( \Lcal_{X_J}  R^{ab}- 
 X^\si_J \om_\si^{ac} R^{cb})+X^\rho_I  \omtilde^1_\rho ( \Lcal_{X_J}\Rtilde^{ab}-
  X^\si_J \om_\si^{ac}  \Rtilde^{cb}) \nonumber \\
 & & ~ ~ ~ ~ ~ ~ ~ ~ ~ ~   -  i_{X_I} ( R^{ab} + \Rtilde^{ab} ) i_{X_J}  R^1 -   i_{X_I} ( R^1 +  \Rtilde^1 ) i_{X_J} R^{ab}] \nonumber\\
    & & ~ ~ ~ ~ ~ ~ ~ ~ ~ ~ + {1 \over 32}  \theta^{IJ}   \theta^{KL} [X_K^\lambda \omtilde^{ab}_\lambda \Lcal_{X_I}
    (X^\tau_L \omtilde^{cd}_\tau) \Lcal_{X_J} R^{cd} - X_K^\lambda \om^{ab}_\lambda  \Lcal_{X_I}
   (X^\tau_L  \om^{cd}_\tau)  \Lcal_{X_J} R^{cd} \nonumber \\
    & & ~ ~ ~ ~ ~ ~ ~ ~ ~ ~ - 4 \Lcal_{X_I} ( X^\lambda_K \om^{ca}_\lambda) \Lcal_{X_J} (2 \Lcal_{X_L} R^{cb} - 2 X^\tau_L \om_\tau^{cd}  R^{db} )+ 8 \Lcal_{X_I} (i_{X_K} R^{ca} )  \Lcal_{X_J} ( i_{X_L} R^{cb })] \nonumber \\ & & \\
   & & R^2= 0 \\
    & & \Rtilde^2 = 0
\ena
~ ~ ~ ~ ~ ~ ~ ~ ~ ~ with $\omtilde^{ab} \equiv \unmezzo \epsilon^{abcd} \om^{cd}$, 
$\Rtilde^{ab}\equiv \unmezzo \epsilon^{abcd} R^{cd}$.
\sk
\noi {\bf Fermion field}
\eqa
\psi^1\!\!\!&=&\!\!\! {1 \over 8} \theta^{IJ} X_I^\mu \om^{ab}_\mu (\ga_{ab} \Lcal_{X_J} \psi + X^\nu_J \om^{ac}_\nu \ga_{bc} \psi) \\
\psi^2\!\!\! &=& \!\!\!- {1\over 8} \theta^{IJ} X^\rho_I  \Big[( \om^1_\rho - i \ga_5
\omtilde^1_\rho)(2 \Lcal_{X_J} \psi- \unquarto X^\si_J \om_\si^{cd}
\ga_{cd} \psi) 
 \nonumber \\
& &~~~~~~~~~~~~~-{i\over 4} X^\rho_I
  \om^{ab}_\rho \ga_{ab}(2 \Lcal_{X_J} \psi^1 - \unquarto X^\si_J
  \om_\si^{cd} \ga_{cd} \psi^1) 
- X^\si_J  ( i \om^1_\si + \ga_5 \omtilde^1_\si ) \psi \Big] \nonumber \\
 & &\!\!\! + {1 \over 64} \theta^{IJ} \theta^{KL} \ga_{ab} [ {1 \over 4} X^\rho_I \om^{ab}_\rho \ga_{cd} \Lcal_{X_K} (X^\si_J \om^{cd}_\si ) \Lcal_{X_L} \psi
    +  \Lcal_{X_K} (X^\rho_I \om^{ab}_\rho ) \Lcal_{X_L}  (2 \Lcal_{X_J} \psi - \unquarto X^\si_J \om_\si^{cd} \ga_{cd} \psi)] \nonumber \\
   \ena

\sect{Compatibility of  SW map and  NC gravity action with the hermiticity and charge conjugation conditions}

Since the hermiticity properties (\ref{hermconjNC}) are essential to ensure reality of
the NC action, it is necessary to check whether the SW solutions indeed satisfy these properties.
This can be seen directly on the SW solutions (\ref{SWsolV}) -
(\ref{SWsolR}), and is due to their gamma matrix structure
 (odd gamma matrices  for $V$, even gamma matrices  for
$\Om$ and $R$) and to the reality of the component fields.

Similarly one can argue for the charge conjugation conditions (\ref{ccc}): the matrix structure of
(\ref{SWsolV}) - (\ref{SWsolR}) indeed is such that these conditions hold, because even terms in $\theta$ multiply symmetric gamma matrices (in the sense that $C\ga$ is symmetric), while odd terms multiply antisymmetric gamma matrices. 

The compatibility between the SW map and the hermiticity and charge conjugation conditions on 
all the fields is proven to all orders  in $\theta$ and for a general field representation  in Appendix B.
\sk
When the noncommutative fields are expressed in terms of the
commutative ones via the SW map,  these latter are considered to be
the elementary dynamical fields. The charge conjugation operation is
then the usual operation on commutative fields, and is extended to the
noncommutativity parameter $\theta$ via the rule
\eq
\theta\to\theta^C=-\theta~
\en
corresponding to  $~\star_\theta\to\star_\theta^C=\star_{-\theta}$, cf. (\ref{defCconj}).
The compatibility of SW map with the charge conjugation condition
reads (see also (\ref{B41}) and (\ref{B42})):
\eq\label{defCconjSW}
\widehat\psi\to{\,\widehat\psi}^{\;C}
=-\gamma_0 C {\,\widehat{\psi}}^{\,\ast}\!~,~~
\widehat{V}\to {\,\widehat{V}}^{\;C}=
C{\,\widehat{V}}^{\;T}\!C\!~,~~{\,\widehat{\Omega}}\to {\,\widehat{\Omega}}^{\;C}=
C{\:\!
\widehat{\Omega}_{}}^{\;T}\!C\!
\!~,
\en
where 
$$\widehat{\psi}=\widehat{\psi}(\psi,\Omega,\theta)~,~~
\widehat{V}=\widehat{V}(V,\Omega,\theta)~,~~
\widehat{\Omega}=\widehat{\Omega} (\Omega,\theta)\;.$$
and 
\eq
{\,\widehat{\psi}}^{\;C}=\widehat{\psi} (\psi^C,\Omega^C,\theta^C)~,~~
{\,\widehat{V}}^{\;C}=\widehat{V}(V^C,\Omega^C,\theta^C)~,~~
{\,\widehat{\Omega}}^{\;C}=\widehat{\Omega} (\Omega^C,\theta^C)
\en
Since the transformations (\ref{defCconjSW}) are the same as
(\ref{defCconj}), they
immediately imply that the noncommutative gravity action
coupled to spinor fields and expanded via SW map  to any order in the 
noncommutativity parameter $\theta$ in terms of the commutative field, is charge conjugation invariant.
As in Section 2.6 this implies that the bosonic action is even in
$\theta$ and that  the fermionic part is also even in $\theta$ if
it describes a Majorana fermion coupled to gravity. 

In particular, the bosonic action must vanish at first order in 
$\theta$, as we verify explicitly in the next Section.

\sect{The noncommutative action expanded to second order in $\theta$}

\subsection{The bosonic action at first order vanishes}

This can be  explicitly verified: let us consider the first order pure gravity action:
\eq
 S^1 \equiv  \int Tr \left(i {R}\westar V \westar V \ga_5  \right)^1 = 
  \int Tr \left(i {R^1}\we V^0 \we V^0 \ga_5  +  i R^0 \we (V \westar V \ga_5)^1 \right) 
 \label{actiongrav1}
\en
\noi where 
\eqa
  & &  (V \westar V \ga_5)^1 = V^1 \we V^0 \ga_5+ V^0 \we V^1 \ga_5+ {i\over 2} \theta^{\rho\si} \part_{\rho} V^0_{\mu} \part_{\si} V^0_{\nu} \ga_5dx^\mu \we dx^\nu  = \nonumber \\
   & &  ~ ~ ~ ~=  \Vtilde^{1~a} \we V^b \ga_a \ga_5 \ga_b + V^b \we \Vtilde^{1~a}  \ga_b \ga_a \ga_5 
     + {i\over 2} \theta^{\rho\si} \part_{\rho} V^a_{\mu} \part_{\si} V^b_{\nu} \ga_a \ga_b \ga_5dx^\mu \we dx^\nu \nonumber \\ & &~ ~ ~ ~= -2 \Vtilde^{1~a} \we V^a \ga_5 + 
    i \theta^{\rho\si} \part_{\rho} V^a_{\mu} \part_{\si} V^a_{\nu} \ga_5dx^\mu \we dx^\nu   
     \ena

and we have used that up to boundary terms:
\eq
 \int Tr \left(i {R}\westar V \westar V \ga_5  \right) =  \int Tr \left(i {R}\we ( V \westar V \ga_5)  \right) 
 \en
Recalling that $R^1$ has only $1$ and $\ga_5$ parts, and $R^0$ has only the $\ga_{ab}$ part, the trace
in (\ref{actiongrav1}) is only over $\ga_{ab}$ or $\ga_{ab} \ga_5$ yielding always $0$. We have
thus verified that the first order part of the pure gravity NC action vanishes.

\subsection{The action at second order}

The second order part of the NC action reads:
\eqa
& & S^2 \equiv  \int Tr \left(i {R}\westar V \westar V \ga_5  
 - (D\psi \star \psibar - \psi \star D\psibar ) \westar V \westar V \westar V
\ga_5 \right)^2 = \nonumber \\
& & ~ ~ ~ ~ = \int Tr [i {R^2}\we V^0 \we V^0 \ga_5  
+ {R^0}\we (V \westar V \ga_5)^2 + R^1 \we (V \westar V \ga_5)^1 \nonumber \\
& & 
 ~ ~ ~ ~  - (D\psi \star \psibar - \psi \star D\psibar )^2 \we V \we V \we V
\ga_5  - (D\psi \star \psibar - \psi \star D\psibar )^0 \we (V \westar V \westar V
\ga_5 )^2  \nonumber \\
& & ~ ~ ~ ~  - (D\psi \star \psibar - \psi \star D\psibar )^1 \we (V \westar V \westar V
\ga_5)^1 ] 
\ena
Expanding the fields as in preceding Section, and carrying out the traces yields:
 \eqa
 & &  S^2  =  \int  (R^{2~ab} \we V^c \we V^d + 2 R^{ab} \we V^{2~c} \we V^d - 2 R^{ab} \we 
  V^{1~c}  \we V^{1~d} ) \epsilon_{abcd} \nonumber \\
  & & ~ ~ ~ ~ ~ ~ ~ - 2 \theta^{IJ} R^{ab} \we \Lcal_{X_I}  V^{1~a} \we \Lcal_{X_J}  V^b + 8 R^1 \we V^{1~a} \we V^a 
  - 4 \theta^{IJ} \Rtilde^1 \we \Lcal_{X_I}  V^a \we \Lcal_{X_J} V^a \nonumber \\
 \ena
 \noi for the pure gravity part, and 
  \eqa
 & &  S^2_\psi = \int [\psibar^2 \ga_{abc} \ga_5 D \psi + \psibar \ga_{abc} \ga_5 (D\psi)^2 + \psibar^1 \ga_{abc} \ga_5 (D\psi)^1 \nonumber \\ 
 & & ~~~~~~~~ +{i\over2} \theta^{IJ} (\Lcal_{X_I}  \psibar^1 \ga_{abc} \ga_5 \Lcal_{X_J}  D\psi + 
 \Lcal_{X_I}  \psibar \ga_{abc} \ga_5 \Lcal_{X_J}  (D\psi)^1) \nonumber \\
  & & ~~~~~~~~ +{1\over 8} \theta^{IJ} \theta^{KL} ((\Lcal_{X_I} \Lcal_{X_K}   \psibar^1) \ga_{abc} \ga_5 \Lcal_{X_J} \Lcal_{X_L}  D \psi+ (\Lcal_{X_I} \Lcal_{X_K} \psibar ) \ga_{abc} \ga_5 \Lcal_{X_J} \Lcal_{X_L}  (D \psi)^1) ] \nonumber \\
     & &~~~~~~~~ \we V^a \we V^b \we V^c  \nonumber\\ 
  & &~~~~~~~~ + [ \psibar^1 \ga_a\ga_b\ga_c D \psi + \psibar \ga_a\ga_b\ga_c (D \psi)^1
  + {i \over 2} \theta^{IJ}\Lcal_{X_I}  \psibar  \ga_a\ga_b\ga_c  \Lcal_{X_J}  D \psi]  \we \nonumber \\
  & &~~~~~~~~ 
   \we ( V^{1~a}  \we V^b \we V^c - V^a \we  V^{1~b} \we V^c + V^a \we V^b \we  V^{1~c} )\nonumber \\
   & &~~~~~~~~ + {i\over 2} \theta^{IJ} [ \psibar^1 \ga_a\ga_b\ga_c \ga_5 D \psi + \psibar \ga_a\ga_b\ga_c \ga_5 (D \psi)^1
  + {i \over 2} \theta^{KL} \Lcal_{X_K}  \psibar  \ga_a\ga_b\ga_c  \ga_5 \Lcal_{X_L}  D \psi]  \we \nonumber \\
  & &~~~~~~~~ 
   \we ( \Lcal_{X_I}  V^a \we \Lcal_{X_J}  V^b \we V^c + V^a \we  \Lcal_{X_I}  V^b \we \Lcal_{X_J}  V^c + \Lcal_{X_I} V^a \we V^b \we  \Lcal_{X_J} V^c )\nonumber \\
   & &~~~~~~~~ + \psibar \ga_a\ga_b\ga_c D\psi \we K^{2~abc} + \psibar  \ga_a\ga_b\ga_c  \ga_5 D\psi \we L^{2~abc} \nonumber \\
    & &~~~~~~~~ - (\psi \leftrightarrow D \psi) 
  \ena
  \noi for the fermi field part, where $K^{2~abc}$ and $L^{2~abc}$ are vielbein combinations originating from  $ (V \westar V \westar V\ga_5 )^2$:
  \eqa
 K^{2~abc} &&\!\!\!\!\!\!\equiv  {i \over 2} \theta^{IJ} ( \Lcal_{X_I} V^{1~a}  \we \Lcal_{X_J}  V^b \we V^c + 
  \Lcal_{X_I}  V^{1~a}  \we V^b \we \Lcal_{X_J}  V^c +  V^{1~a}  \we \Lcal_{X_I}  V^b \we  \Lcal_{X_J}  V^c \nonumber\\
   & &~+ \Lcal_{X_I}  V^{a}  \we \Lcal_{X_J}  V^{1~b} \we V^c + 
  \Lcal_{X_I}  V^{a}  \we V^{1~b} \we \Lcal_{X_J}  V^c +  V^{a}  \we \Lcal_{X_I}  V^{1~b}\we  \Lcal_{X_J} V^c \nonumber\\
  & &~ +\Lcal_{X_I}  V^{a}  \we \Lcal_{X_J} V^b \we V^{1~c} + 
  \Lcal_{X_I}  V^{a}  \we V^b \we \Lcal_{X_J}  V^{1~c}+  V^{a}  \we \Lcal_{X_I}  V^b \we  \Lcal_{X_J}  V^{1~c} ) \nonumber \\
  \ena
  \eqa
    L^{2~abc} &&\!\!\!\!\!\! \equiv V^{2~a}  \we V^b \we V^c - V^{1~a}  \we V^{1~b} \we V^c + V^a \we V^{2~b} \we V^c      
   \nonumber \\
   & &  + V^{a}  \we V^b \we V^{2~c}  + V^{1~a}  \we V^{b} \we V^{1~c}  - V^a \we V^{1~b} 
   \we V^{1~c}      
   \nonumber \\
      & &  - \unquarto \theta^{IJ} \theta^{KL} (\Lcal_{X_I}  \Lcal_{X_K}  V^a \we \Lcal_{X_L}  V^b \we \Lcal_{X_J}  V^c
    + \Lcal_{X_K}  V^v \we \Lcal_{X_I}  \Lcal_{X_L}  V^b \we \Lcal_{X_J}  V^c )\nonumber \\
    & &  - {1 \over 8}  \theta^{IJ} \theta^{KL}   \Lcal_{X_I}  (\Lcal_{X_K} V^a \we V^b
    + V^a \we \Lcal_{X_K}  V^b)  \we \Lcal_{X_J}  \Lcal_{X_L}  V^c 
    \ena
{\bf Note:}
Lie derivatives in the action act on forms in a  diffeomorphic
invariant way. Hence the action is diffeomorphisms invariant.
However, the action of $ \Lcal_{X_I}$ on a Lorentz tensor is {\sl not}
Lorentz covariant, since for example,  using the spin connection
$\omega^{ab}$ with vanishing torsion,
\eq
 \Lcal_{X_I} V^a= D{X_I^a} + (i_{X_I} \om^{ab})V^b\nonumber \\
\en
with $X^a_I=X_I^\mu V_\mu^a$, $DX^a_I=dX^a_I-\omega^{ab}X_I^b$.
This expression explicitly
contains a non-Lorentz covariant term due to the ``naked'' connection
$\omega^{ab}$. Only with successive integrations by parts one recovers
the manifest local Lorentz invariance of the action (that we recall is
guaranteed by invariance of the noncommutative action under
noncommutative local Lorentz transformations and by the SW map construction).
Thus, if written in terms of the form components, only {\it  bona
  fide} Lorentz  tensors appear in the action, with all indices contracted to yield a scalar.

\sect{Conclusions}

The fields and the action of the NC vielbein gravity (+ fermions) constructed in
\cite{AC1} have been expanded via the SW map to second order in the noncommutativity
parameter $\theta$.
The expanded action 
involves only the classical (commuting) fields of usual gravity coupled to fermions, and
the background commuting vector fields $X_I$ that define the abelian twist.
The action is {\it real}, thanks to the compatibility of the SW map with the hermiticity
conditions on the field; it is also charge
conjugation invariant due to the compatibility of the SW map with
the charge conjugation conditions on the fields. This implies that the bosonic
action is {\it even in} $\theta$.  In Appendix B these compatibilities are shown in general, without reference to the specific model considered in this paper.

The expanded action is invariant under usual diffeomorphisms and local Lorentz transformations.

In its use in a geometric theory, we found convenient to reformulate
the SW map in the geometric language of exterior forms: this allowed to
generalize the SW map to arbitrary abelian twists.

\section*{Acknowledgements}
We thank Marija Dimitrijevic, Giampiero Esposito and Elisabetta Di Grezia for useful
conversations on the subject.

 \appendix

\sect{Twist differential geometry}

The noncommutative deformation of gravity considered here and in ref. \cite{AC1}
relies on the existence (in the deformation quantization context, see
for ex \cite{book} ) of an associative $\star$-product between
functions and more generally an associative $\westar$ exterior product between forms that
satisfies the following properties:
\sk
\noi
$\bullet~~$ \noi Compatibility with the undeformed exterior differential:
\eq
d(\tau\wedge_\star \tau')=d(\tau)\wedge_\star \tau'=\tau\wedge_\star
d\tau'
\en
$\bullet~~$ Compatibility with the undeformed integral (graded cyclicity property):
        \eq
       \int \tau \westar \tau' =  (-1)^{deg(\tau) deg(\tau')}\int \tau' \westar \tau\label{cycltt'}
       \en
      \noi with $deg(\tau) + deg(\tau')=$D=dimension of the spacetime
      manifold, and where here $\tau$ and $\tau'$ have compact support
      (otherwise stated we require (\ref{cycltt'}) to hold up to
      boundary terms).
\sk
\noi $\bullet~~$ Compatibility with the undeformed complex conjugation:
\eq
       (\tau \westar \tau')^* =   (-1)^{deg(\tau) deg(\tau')} \tau'^* \westar \tau^*
\en
{}Following \cite{AC1}  we describe here a (quite wide) class of twists whose   $\star$-products
 have all these properties.
As a particular case we
have the Groenewold-Moyal $\star$-product
\begin{equation}
f\star g = \mu \big{\{} e^{\frac{i}{2}\theta^{\rho\sigma}\partial_\rho \otimes\partial_\sigma}
f\otimes g \big{\}} , \label{MWstar}
\end{equation}
where the map $\mu$  is the usual pointwise
multiplication: $\mu (f \otimes g)= fg$, and $\theta^{\rho\sigma}$ is a constant
antisymmetric matrix.

\sk

\noi{\bf Abelian Twist}
\sk
\noi Let $\Xi$ be the linear space of smooth vector fields on a smooth manifold $M$, and $U\Xi$ its
universal enveloping algebra. A twist  ${\cal F} \in U\Xi \otimes U\Xi$
defines the associative $\star$-product
\begin{eqnarray}
f\star g &=& \mu \big{\{} {\cal F}^{-1} f\otimes g \big{\}}
\end{eqnarray}
\noi  where the map $\mu$  is the usual pointwise
multiplication: $\mu (f \otimes g)= fg$. The product associativity relies on the defining properties of the twist \cite{Wessgroup,book}.

   \noi Explicit examples of twist are provided by the so-called abelian twists:
\eq
{\cal F}^{-1}= e^{\frac{i}{2}\theta^{IJ}X_I \otimes X_J} \label{Abeliantwist}
\en
where $\{X_I\}$ is a set of mutually commuting vector fields globally
defined on the manifold, and $\theta^{IJ}$ is a constant
antisymmetric matrix. The corresponding $\star$-product is in general
position dependent because the vector fields $X_a$ are in general
$x$-dependent. In the special case that there exists a
global coordinate system on the manifold we can consider the
vector fields $X_a={\partial \over \partial x^a}$. In this instance we have
the Moyal twist, cf. (\ref{MWstar}):
  \eq
   {\cal F}^{-1}=  e^{\frac{i}{2}\theta^{\rho\sigma}\partial_\rho \otimes\partial_\sigma} \label{Mtwist}
   \en

  \noi {\bf Deformed exterior product}
    \sk

   \noi For abelian twists (\ref{Abeliantwist}), the deformed exterior product between forms is defined as
   \eqa
   & & \tau \westar \tau' \equiv \sum_{n=0}^\infty \left({i \over 2}\right)^n \theta^{I_1J_1} \cdots \theta^{I_nJ_n}
   (\Lcal_{X_{I_1}} \cdots \Lcal_{X_{I_n}} \tau) \we  (\Lcal_{X_{J_1}} \cdots \Lcal_{X_{J_n}} \tau')  \nonumber \\
  & & ~~ = \tau \we \tau' + {i \over 2} \theta^{IJ} (\Lcal_{X_I} \tau) \we (\Lcal_{X_J} \tau') + {1 \over 2!}  {\left( i \over 2 \right)^2} \theta^{I_1J_1} \theta^{I_2J_2}  (\Lcal_{X_{I_1}} \Lcal_{X_{I_2}} \tau) \we
 (\Lcal_{X_{J_1}} \Lcal_{X_{J_2}} \tau') + \cdots \nonumber 
  \label{defwestar}
  \ena
       \noi where the commuting tangent vectors $X_I$ act on forms via the Lie derivatives
       ${\cal L}_{X_I} $. 
     This product is associative, and the above formula holds also for $\tau$ or $\tau'$ being a $0$-form (i.e. a function). 
 
\sk
\noi {\bf Exterior derivative}
        \sk
         \noi The exterior derivative satisfies the usual (graded) Leibniz rule,
         since it commutes with the Lie derivative:
        \eqa
        & & d (f \star g) = df \star g + f \star dg \\
        & & d(\tau \westar \tau') = d\tau \westar \tau'  + (-1)^{deg(\tau)} ~\tau \westar d\tau'
        \ena

\sk

       \noi {\bf Integration: graded cyclicity} \nopagebreak
        \sk
        \noi If we consider an abelian twist (\ref{Abeliantwist})
        given by globally defined commuting vector fields $X_a$,
        then the usual integral is cyclic under the $\star$-exterior
        products of forms, i.e., up to boundary terms,
        \eq
       \int \tau \westar \tau' =  (-1)^{deg(\tau) deg(\tau')}\int \tau' \westar \tau
       \en
      \noi with $deg(\tau) + deg(\tau')\!=\,$D$\,$= dimension of the spacetime
      manifold. In fact we have, up to boundary terms,
\eq       \int \tau \westar \tau' =    \int \tau \wedge \tau'=
(-1)^{deg(\tau) deg(\tau')}\int \tau' \wedge \tau=
(-1)^{deg(\tau) deg(\tau')}\int \tau' \westar \tau
\en
For example at first order in $\theta$,
\eq
\int \tau \westar \tau' =    \int \tau \wedge \tau'-{i\over
 2}\theta^{ab}\int{\cal L}_{X_a}(\tau\wedge {\cal L}_{X_b}\tau')
=
\int \tau \wedge \tau'-{i\over
 2}\theta^{ab}\int d {i}_{X_a}(\tau\wedge {\cal L}_{X_b}\tau')
\en
where we used the Cartan formula ${\cal L}_{X_a}=di_{X_a}+i_{X_a}d$.
\sk       \noi {\bf Complex conjugation}
    \sk
        \noi If we choose real fields $X_a$ in the definition of the
        twist (\ref{Abeliantwist}),  it is immediate to verify that:
        \eq
        (f \star g)^* = g^* \star f^*\label{starfg*}
        \en
        \eq
        (\tau \westar \tau')^* =   (-1)^{deg(\tau) deg(\tau')} \tau'^* \westar \tau^*\label{startt*}
        \en
        since sending $i$ into $-i$ in the twist (\ref{Mtwist}) amounts to send $\theta^{ab}$ into
        $-\theta^{ab} = \theta^{ba}$, i.e. to exchange the
        order of the factors in the $\star$-product.

        \sk

\section{Seiberg Witten map: hermiticity and charge conjugation properties}
We here  sharpen the results of \cite{AJSW} concerning the general
properties of hermiticity and reality of SW map (in particular in \cite{AJSW} the
gauge group was an internal group, here it can also be the Lorentz
group, for example in its spin representation).  Applying these
general properties we derive the hermiticity and 
charge conjugation properties of the fields of NC gravity coupled 
to spinors used in Section 2.5. 

\sk
Given a representation $\rho:G\rightarrow GL(n,
{C})$ of a
group $G$ we can consider three
other representations of $G$:  
i) the inverse Hermitian representation, defined for all $g\in G$ by
$
\rho'(g)={\rho(g)^{-1}}^\dagger
$
(without inversion we would not have $\rho'(g\tilde
g)=\rho'(g)\rho'(\tilde g)$); ii) the complex conjugate representation,
$
\rho^\ast(g)=\rho(g)^\ast
$
where ${}^\ast$ denotes complex conjugation, iii) the inverse
transpose representation $\rho^T(g)={\rho(g)^{-1}}^T$ (i.e.  the complex conjugate representation of $\rho'$). 
The representation $\rho$ is unitary if $\rho'=\rho$.

These representations induce representations of the Lie algebra 
${\mathtt g}=\,$Lie$(G)$. If $g=e^{i\lambda_aT^a}$ with $\lambda_a\in 
{R}$,
we find 
\eq
\rho'(T^a)=[\rho(T^a)]^\dagger~,~~\rho^\ast(T^a)=-[\rho(T^a)]^\ast~\label{defrhoastrho'},~~
\rho^T(T^a)=-[\rho(T^a)]^T~.
\en In turn these Lie algebra representations can be extended to
representations of the universal enveloping algebra $U{\mathtt g}$ of
$\mathtt{g}$ by linearity and multiplicativity  (we recall that $U{ \mathtt g}$
is the associative algebra of polynomials of elements of $\mathtt g$,
where  the commutator $TT'-T'T$ is identified with the Lie bracket $[T,T']$).
\sk
{}From (\ref{defrhoastrho'}) we have
$$\rho'(A)=\rho(A)^\dagger~,~~\rho^\ast(A)=-\rho(A)^\ast~, \rho^T(A)=-\rho(A)^T~.$$
We show that these relations hold also for the NC fields.  In
other words the  SW map is compatible with 
hermitian conjugation, with complex conjugation and with transposition:
\eq
\widehat{\rho'(A)}={\widehat{\rho(A)\,}}^\dagger~,~~
\widehat{\rho'(\L)}={\widehat{\rho(\L)\,}}^\dagger \label{her}
\en
i.e., $\widehat{(\rho (A))^{{^{{\!}}}\dagger}{}~{}}\!={\widehat{\rho(A)\,}}^\dagger\;,~
\widehat{(\rho (\L))^{{^{{\!}}}\dagger}{}~{}}\!={\widehat{\rho(\L)\,}}^\dagger{}\,,
$
that for short we rewrite $\widehat{A^\dagger}=
{\widehat A}^{^{{{\,{\scriptstyle{\dagger}}}}}}{}\,,~
\widehat{\L^\dagger}={\widehat\L}^{^{{{\,{\scriptstyle{\dagger}}}}}}{}\,$, 
\eq
\widecheck{\rho^\ast (A)}=-{{\widehat{\rho(A)\,}}}^\ast~~,~~~
\widecheck{\rho^\ast(\L)}=-{{\widehat{\rho(\L)\,}}}^\ast~,
\label{conj}\en
and
\eq
\widecheck{\rho^T (A)}=-{{\widehat{\rho(A)\,}}}^T~~,~~~
\widecheck{\rho^T(\L)}=-{{\widehat{\rho(\L)\,}}}^T~,
\label{transp}\en
where $\widecheck{~{}~{}~{}~}$ denotes the SW map
with $-\theta$ noncommutativity. More explicitly 
formula  (\ref{transp}) reads
$$SW[\rho^T(A),-\theta]=-SW[\rho(A),\theta]^T ~,~~
SW[\rho^T(\epsi)\rho^T(A),-\theta]=-SW[\rho(\epsi),\rho(A),\theta]^T ~;
$$
and similarly for formulae (\ref{conj}) and (\ref{her}).  
\sk\noi
{\it Proof of  (\ref{her}), (\ref{conj}) and (\ref{transp}). }
We recall that for generic space-time 
dependent matrices $M$ and $N$, under complex conjugation, 
transposition and hermitian conjugation we have
\eq
{(M\star N)}^\ast={M}^\ast \star_{-\theta} {N}^\ast
{}~~~~,~~~~~~~(M\*N)^T=N^T\star_{-\theta} M^T
{}~~~~,~~~~~~~(M\*N)^\dagger=N^\dagger\*M^\dagger~~.
\label{mn}
\en
The Hermitian conjugates of the relations  (\ref{rone}),
(\ref{rtwo}) in the representation
$\rho$, are
    \eqa
   & &
   {\widehat{\rho(A_\mu)}}^{n+1\;\dagger}= -{1 \over 4(n+1)} \theta^{\rho\sigma} \{\widehat{\rho(A_\rho)}^\dagger, \part_\sigma \widehat{\rho(A_\mu)}^\dagger + \widehat{\rho(F_{\sigma\mu})}^\dagger \}^n_\star \\
    & & {{{\widehat{\rho(\epsi)}}^{n+1\;\dagger}}}=  -{1 \over 4(n+1)} \theta^{\rho\sigma} \{\widehat{\rho(A_\rho)}^\dagger, \part_\sigma \widehat{\rho(\epsi)}^\dagger \}^n_\star~;
     \ena
where by definition $
   \widehat{\rho(F_{\nu\rho})} \equiv \part_\nu \widehat{(\rho(A_\rho)} -  \part_\rho \widehat{\rho(A_\nu)} - i \widehat{\rho(A_\nu)} \star \widehat{\rho(A_\rho)}
    + i \widehat{\rho(A_\rho)} \star \widehat{\rho(A_\nu)} .
$
 Relations  (\ref{rone}), (\ref{rtwo}) in the representation
$\rho'$, are
   \eqa
   & &
   {\widehat{\rho'(A_\mu)}}^{n+1}= -{1 \over 4(n+1)} \theta^{\rho\sigma} \{\widehat{\rho'(A_\rho)}, \part_\sigma \widehat{\rho'(A_\mu)} + \widehat{\rho'(F_{\sigma\mu})} \}^n_\star \\
    & & {{{\widehat{\rho'(\epsi)}}^{n+1}}}= -{1 \over 4(n+1)} \theta^{\rho\sigma} \{\widehat{\rho'({A_\rho)}}, \part_\sigma \widehat{\rho'(\epsi)} \}^n_\star~.
     \ena
Since these two sets of relations have the same structure in terms of
their respective variables 
${\widehat{\rho(A)}}^\dagger, \widehat{\rho(\L)}^\dagger$ and ${\widehat{\rho'(A)}}, \widehat{\rho'(\L)}$, and
since at zeroth order in $\theta$,
$\rho(A)^\dagger=\rho'(A)$, $\rho(\L)^\dagger=\rho'(\L)$ the compatibility (\ref{her}) 
is iteratively proven at all orders in $\theta$.

\sk
We proceed similarly in order to prove (\ref{conj}); the complex
conjugate of the relations  (\ref{rone}),
(\ref{rtwo}) in the representation
$\rho$ can be written as
    \eqa
   & &
   -{\widehat{\rho(A_\mu)}}^{n+1\;\ast}= {1 \over 4(n+1)} \theta^{\rho\sigma} \{-\widehat{\rho(A_\rho)}^\ast, -\part_\sigma \widehat{\rho(A_\mu)}^\ast - \widehat{\rho(F_{\sigma\mu})}^\ast \}^n_{\star_{-\theta}} \\
    & & -{{{\widehat{\rho(\epsi)}}^{n+1\;\ast}}}=  {1 \over 4(n+1)} \theta^{\rho\sigma} \{-\widehat{\rho(A_\rho)}^\ast, -\part_\sigma \widehat{\rho(\epsi)}^\ast \}^n_{\star_{-\theta}}~.
     \ena
Relations  (\ref{rone}), (\ref{rtwo}) in the representation
$\rho^\ast$ and with noncommutativity parameter $-\theta$ (and
corresponding star product $\star_{-\theta}$) read
    \eqa
   & &
   {\widecheck{\rho^\ast(A_\mu)}}^{n+1}= {1 \over 4(n+1)} \theta^{\rho\sigma} \{\widecheck{\rho^\ast(A_\rho)}, \part_\sigma \widecheck{\rho^\ast(A_\mu)} + \widecheck{\rho^\ast(F_{\sigma\mu})} \}^n_{\star_{-\theta}} \\
    & & {{{\widecheck{\rho^\ast(\epsi)}}^{n+1}}}=  {1 \over 4(n+1)} \theta^{\rho\sigma} \{\widecheck{\rho^\ast(A_\rho)}, \part_\sigma \widecheck{\rho^\ast(\epsi)} \}^n_{\star_{-\theta}}~.
     \ena
Since these two sets of relations have the same structure in terms of
their respective variables 
$-{\widehat{\rho(A)}}^\ast, -\widehat{\rho(\L)}^\ast$ and ${\widecheck{\rho^\ast(A)}}, \widecheck{\rho^\ast(\L)}$, and 
since at zeroth order in $\theta$,
$-\rho(A)^\ast=\rho^\ast(A)$, $-\rho(\L)^\ast=\rho^\ast(\L)$, the compatibility (\ref{her}) 
is iteratively proven at all orders in $\theta$.

{}From (\ref{her}) and (\ref{conj}) easily follows the
compatibility of the SW map with transposition, eq. (\ref{transp}).
\sk
A similar iterative procedure shows that for matter fields
transforming in the adjoint (see (\ref{phiadrec}) and
(\ref{phiadrecgen})), given $\rho'(\phi)=\rho(\phi)^\dagger$, $\rho^\ast(\phi)=-\rho(\phi)^\ast$ and $\rho^T(\phi)=-\rho(\phi)^T$,  we have
$SW[\rho'(\phi),\rho'(A),\theta]=SW[\rho(\phi),\rho(A),\theta]^{\,\dagger}$
and
$$SW[\rho^\ast(\phi),\rho^\ast(A),-\theta]=-SW[\rho(\phi),\rho(A),\theta]^{\,\ast}~,~~
SW[\rho^T(\phi),\rho^T(A),-\theta]=-SW[\rho(\phi),\rho(A),\theta]^{\,T},$$
that we rewrite as
\eq
\widehat{\rho(\phi)'} =\widehat{\rho(\phi)}^{\,\dagger}
~,~~ \widecheck{\rho^\ast(\phi)} =-\widehat{\rho(\phi)}^{\,\ast}~,~~
\widecheck{\rho^T(\phi)} =-\widehat{\rho(\phi)}^{\,T}~;\label{Vdaggertransp}
\en
with abuse of notations we  simply write the first of these relations  $\widehat{\,\phi^\dagger\,}
=\widehat{\phi}^{\;\dagger}$.  

Finally the same iterative procedure shows that for matter field in
the fundamental (see (\ref{phirec}) and (\ref{phirecgen})), given
$\rho^\ast(\phi)=\rho(\phi)^\ast$, we have
$SW[\rho^\ast(\phi),\rho^\ast(A),-\theta]=SW[\rho(\phi),\rho(A),\theta]^{\,\ast}$, i.e.,
\eq\label{SWpgiast}
\widecheck{\,\phi^\ast} ={\,\widehat{\phi}}^{\:\ast}~.
\en
Charge conjugation is the transformation that maps a particle representation of a
symmetry group to its complex conjugate representation ($\rho\to\rho^\ast$), hence the
compatibility of SW map with complex conjugation is the compatibility
with charge conjugation. On spinor fields complex conjugation and
charge conjugation differ by a unitary transformation (see
(\ref{CCCOF}), 
(\ref{unitarytr})). As we show below also
in this case the SW map is compatibe with charge conjugtion.

\sk

\noi {\bf Spin representation of Lorentz group}\\
We now consider the gauge group $G=SL(2,C)$ and fix the
representation $\rho$ (that from now on we omit writing) to be the one
determined by the $4\times 4$
matrices in the spinor representation used throughout the paper (the
$(\frac{1}{2},\frac{1}{2})$-sinor representation). We prove that $\widehat{\Omega}=i\widehat{A}$ satisfies the
hermiticity and symmetry conditions,
\eq
\widehat{{}\Omega{}}^{\:\dagger}=-\gamma_0\widehat\Omega\gamma_0~~,~~~\widehat{\Omega}^{\:T}=C\widecheck{
 \Omega } C\,; \label{ohat}
\en
these are precisely the conditions 
(\ref{hermconjNC}) and (\ref{ccc}), that we have here rewritten using
hatted variables in order to stress that these conditions are now
{\sl derived} from the hermiticity and symmetry properties of the classical
fields and of the SW map.

We know that the classical fields satisfy
$\Omega^\dagger=-\gamma_0\Omega\gamma_0\,,~\Omega^{\,T}=C\Omega C$,
i.e., since $A=-i\Omega$, $\gamma_0=\gamma_o^{-1}$, $C=-C^{-1}$, we
know that
$A^{\dagger}=\gamma_0 A\gamma_0^{-1},~-A^{T}=C A C^{-1}$.
We then have
\eq\label{AAAhat}
\widehat{A}^{\,\dagger}=\widehat{A^\dagger}=\widehat{\gamma_0 A
 \gamma_0^{-1}}=\gamma_0\widehat{A}\gamma_0^{-1}~,
\en
\eq
\widehat{A}^{\,T}=-\widecheck{\rho^T(A)}=-\widecheck{(-A^T)}=-\widecheck{~C A
 C^{-1}}=- C\widecheck{\,A\,}C^{-1}~,
\en
where we used (\ref{conj}) and (\ref{transp}), and in the last equality of each expression we used compatibility of SW map with the
similarity transformations $A\to \gamma_0 A \gamma_0^{-1}$ and  $A\to C
A C^{-1}$ respectively (for example $SW[CAC^{-1},\theta]=C\:\!SW[A,\theta]C^{-1}$). These expressions
immedialtely imply (\ref{ohat}).

A similar proof shows that the gauge parameter $\widehat\epsilon$, as
well as the vielbein $\widehat{V}$
satisfy the hermiticity and symmetry conditions,
\eq
\widehat{{}\epsilon{}}^{\:\dagger}=-\gamma_0\widehat\epsilon\gamma_0~~,~~~\widehat{\epsilon}^{\:T}=C\widecheck{\,
\epsilon\,} C\,; \label{ehat}
\en
\eq
\widehat{{}V{}}^{\:\dagger}=\gamma_0\widehat V\gamma_0~~,~~~\widehat{V}^{\:T}=C\widecheck{\,
V\,} C\,; \label{vhat}
\en
hint: use that
$\widehat\epsilon=i_{}\widehat\varepsilon$,
where $\widehat\epsilon$ corresponds to the gauge potential $\widehat\Omega$, and
$\widehat\varepsilon$ corresponds to the gauge potential 
$\widehat A$ (see  (\ref{stargauge}) and (\ref{3.44h})). For
(\ref{vhat}) use (\ref{Vdaggertransp}). Relations (\ref{vhat})
coincide with the reality and charge conjugation conditions 
considered in  (\ref{hermconjNC}) and (\ref{ccc}).
\sk

\noi{\bf Charge Conjugation}\\
Under charge conjugation we have
\eq
\psi\to\psi^{\;C}=C(\bar\psi)^T=-\gamma_0 C \psi^\ast\!~,~~
V\to V^{\,C}=V
~,~~
A\to A^{\,C}=A
\!~\label{CCCOF}
\en
Applying the SW map we obtain the
noncommutative fields relations $\widehat{V^C}=\widehat{V},
\widehat{A^C}=\widehat{A}$, that with the help of (\ref{vhat}) and
(\ref{AAAhat}) equivalently read
\eq\label{B41}
\widecheck{V^\CCC}=C{\,\widehat{V}}^{\,T}{C}~,~~
\widecheck{A^\CCC}=C{\,\widehat{A}}^{\;T}{C}~.
\en
The charge conjugation operation on the gauge potential can also
be written 
\eq A\to A^C=(-\gamma_0C) \,\rho^\ast(A) (-\gamma_0C)^{-1}=-(-\gamma_0C) \,A^\ast (-\gamma_0C)^{-1}\label{unitarytr}~.
\en
Then
\eqa\label{B42}
\widecheck{\psi^\CCC}&=&SW[\psi^C,A^C,-\theta]=SW[-\gamma_0C\psi^\ast,Ad_{-\gamma_0C}\,\rho^\ast(A),-\theta]=-\gamma_0C\,SW[\psi^\ast,\rho^\ast(A),-\theta]\nonumber\\
&=&-\gamma_0C{\,\widecheck{\psi^\ast}}=-\gamma_0C{\,\widehat{\psi}}^{\;\ast}~
\ena
where in the last passage we used (\ref{SWpgiast}). 

In the previous sections
we used the notation $\widehat{\psi}=\widehat\psi(\psi,\Omega,\theta)$
to denote the SW map $\widehat{\psi}=SW[\psi,A,\theta]$,
and similarly for $\widehat{V}$ and
$\widehat\Omega=i\widehat{A}$. Then relations (\ref{B41}) and
(\ref{B42}) coincide with  equations (\ref{defCconjSW}).

\sect{Gamma matrices in $D=4$}

We summarize in this Appendix our gamma matrix conventions in $D=4$.

\eqa
& & \eta_{ab} =(1,-1,-1,-1),~~~\{\ga_a,\ga_b\}=2 \eta_{ab},~~~[\ga_a,\ga_b]=2 \ga_{ab}, \\
& & \ga_5 \equiv i \ga_0\ga_1\ga_2\ga_3,~~~\ga_5 \ga_5 = 1,~~~\epsi_{0123} = - \epsi^{0123}=1, \\
& & \ga_a^\dagger = \ga_0 \ga_a \ga_0, ~~~\ga_5^\dagger = \ga_5 \\
& & \ga_a^T = - C \ga_a C^{-1},~~~\ga_5^T = C \ga_5 C^{-1}, ~~~C^2 =-1,~~~C^\dagger=C^T =-C
\ena

\subsection{Useful identities}

\eqa
 & &\ga_a\ga_b= \ga_{ab}+\eta_{ab}\\
 & & \ga_{ab} \ga_5 = {i \over 2} \epsilon_{abcd} \ga^{cd}\\
 & &\ga_{ab} \ga_c=\eta_{bc} \ga_a - \eta_{ac} \ga_b -i \epsi_{abcd}\ga_5 \ga^d\\
 & &\ga_c \ga_{ab} = \eta_{ac} \ga_b - \eta_{bc} \ga_a -i \epsi_{abcd}\ga_5 \ga^d\\
 & &\ga_a\ga_b\ga_c= \eta_{ab}\ga_c + \eta_{bc} \ga_a - \eta_{ac} \ga_b -i \epsi_{abcd}\ga_5 \ga^d\\
 & &\ga^{ab} \ga_{cd} = -i \epsi^{ab}_{~~cd}\ga_5 - 4 \de^{[a}_{[c} \ga^{b]}_{~~d]} - 2 \de^{ab}_{cd}
 \ena
where
$\delta^{ab}_{cd}=\frac{1}{2}(\delta^a_c\delta^b_d-\delta^b_c\delta^a_d)$
and indices antisymmetrization in square brackets has total weight $1$. 
 \subsection{Charge conjugation and Majorana condition}

\eqa
 & &   {\rm Dirac~ conjugate~~} \psibar \equiv \psi^\dagger
 \ga_0\\
 & &  {\rm Charge~ conjugate~spinor~~} \psi^C = C (\psibar)^T  \\
 & & {\rm Majorana~ spinor~~} \psi^C = \psi~~\Rightarrow \psibar =
 \psi^T C
 \ena

 \vfill\eject

\end{document}